\documentclass[11pt]{article}

\usepackage[english]{babel}
\usepackage[utf8]{inputenc}
\usepackage[usenames,dvipsnames]{color}

\usepackage[a4paper,left=2.5cm,right=2.5cm,top=2.5cm,bottom=2.5cm,
bindingoffset=0mm]{geometry}

\usepackage{cite}

\usepackage{amsthm}
\usepackage{amssymb}
\usepackage{amsmath}
\usepackage{algorithm}
\usepackage[noend]{algpseudocode}

\usepackage{latexsym}

\usepackage{tikz}
\usetikzlibrary{positioning}

\usepackage{graphicx}

\usepackage[onehalfspacing]{setspace}

\begin{document}
\tikzset{%
	every neuron/.style={
		circle,
		draw,
		minimum size=1cm
	},
	neuron missing/.style={
		draw=none, 
		scale=3,
		text height=0.333cm,
		execute at begin node=\color{black}$\vdots$
	},
}
\thispagestyle{empty}
\vspace*{-1.5cm}
\begin{flushright}
  {\small
  LMU-ASC 20/21\\
  MPP-2021-108
  }
\end{flushright}

\vspace{1.75cm}

\begin{center}
{\LARGE
Revealing systematics in phenomenologically viable flux vacua with reinforcement 
learning}
\end{center}

\vspace{0.4cm}

\begin{center}
  Sven Krippendorf$^1$, Rene Kroepsch$^{1}$, Marc Syvaeri$^{1,2}$ 
\end{center}
\vspace{0.3cm}
\begin{center} 
\textit{$^{1}$\hspace{1pt} Arnold Sommerfeld Center for Theoretical 
Physics\\[1pt]
Ludwig-Maximilians-Universit\"at \\[1pt]
Theresienstra\ss e 37 \\[1pt]
80333 M\"unchen, Germany}
\\[1em]
\textit{$^{2}$\hspace{1pt} Max-Planck-Institut f\"ur Physik\\[1pt]
F\"ohringer Ring 6 \\[1pt]
80805   M\"unchen, Germany}
\end{center} 
\vspace{0.8cm}

\begin{abstract}
The organising principles underlying the structure of phenomenologically viable 
string vacua can be 
accessed by sampling such vacua. In many cases this is prohibited by the 
computational cost of 
standard sampling methods in the high dimensional model space. Here we show how 
this problem can 
be alleviated using reinforcement learning techniques to explore string flux 
vacua. We demonstrate in the 
case of the type IIB flux landscape that vacua with requirements on the 
expectation value of the 
superpotential and the string coupling can be sampled significantly faster by 
using reinforcement 
learning than by using metropolis or random sampling.
Our analysis is on conifold and symmetric torus background geometries. 
We show that 
reinforcement learning is able to exploit successful strategies for identifying 
such phenomenologically 
interesting vacua. The strategies are interpretable and reveal previously unknown 
correlations in the flux landscape.

\end{abstract}
\newpage
\tableofcontents

\section{Introduction}
The string theory flux landscape covers a vast range of low-energy effective 
field theories~\cite{Bousso_2000,
Susskind:2003kw,Douglas_2003}. A subset of such theories might be 
phenomenologically desirable, e.g.~featuring 
a weak string coupling $g_s$ and a particular scale of supersymmetry breaking 
$m_{3/2}$, largely determined by 
the expectation value of the flux superpotential $W_0.$\footnote{Work regarding 
properties of the distribution of 
phenomenological properties such as the supersymmetry breaking scale can be 
found 
in~\cite{Ashok_2004,Denef_2004,Douglas_2004,ArkaniHamed:2005yv,Denef_2005,
SUSSKIND_2005,Dine_2008,Conlon_2004,Kallosh_2004,Marchesano_2005,Dine_2005,
Acharya_2005,Dienes_2006,Gmeiner_2006,Douglas_2007_2,Sumitomo:2012wa,
MartinezPedrera:2012rs,Cicoli:2013cha,Broeckel:2020fdz,Broeckel:2021dpz}.}

Apriori, it is unclear whether such phenomenological structures single out a 
particular subset of flux configurations 
and what the common features are in the UV representation. By sampling the space 
of flux vacua such properties 
can be revealed. Although in few parameter models an exploration via random or 
even complete sampling is 
tractable, more realistic models require a different sampling strategy. In fact 
it has been argued that the computational complexity of finding flux vacua is NP 
hard~\cite{Denef:2006ad,Denef:2017cxt,Halverson:2018cio} but phenomenologically 
relevant vacua can nevertheless be found with appropriate optimisation 
techniques~\cite{Bao:2017thx}. More widely speaking advances in artificial 
intelligence techniques make it possible to analyse this problem with a variety 
of approaches (cf.~\cite{Ruehle:2020jrk} for an overview).

Here, we exemplify for the first time how reinforcement learning (RL) can be 
used to search for such correlations in 
the flux landscape (see~\cite{Sutton1998} for a classic textbook on RL). RL has  
previously been utilised to identify efficiently models of particles physics 
from string theory~\cite{Halverson_2019,Larfors:2020ugo,Harvey:2021oue}.

For simplicity and, in addition, to compare with other sampling approaches we 
restrict this 
analysis to few parameter examples which also have been analysed with the help 
of genetic 
algorithms~\cite{Cole:2019enn}. We are able to show that RL can outperform 
metropolis sampling in 
efficiency in this context. Both approaches reveal correlations among flux 
quanta -- previously unreported in the 
literature to our knowledge. Our RL agents however have learned to explore the 
flux environment such that they 
can navigate in nearby solution space, i.e.~to remain in the fundamental domain 
and to keep a particular scale of 
supersymmetry breaking.

The rest of the paper is organised as follows. In 
Section~\ref{sec:fluxenvironments} we describe the flux examples, 
how the RL environment is generated and which reward structure is visible. 
Section~\ref{sec:experiments} contains our 
main analysis of these examples and compares the performance of RL approaches 
with metropolis and 
random searches. In Section~\ref{sec:conclusions} we conclude. Details on the RL 
algorithms we use can be found in Appendix~\ref{app:RL}, our hyperparameter 
searches can be 
found in Appendix~\ref{app:hyperparameters}, and more experiments on the torus background are summarised in Appendix~\ref{app:torusenvironments}.

\section{Flux environments}
\label{sec:fluxenvironments}
At this stage our analysis is based on fixed background geometries. Here we set 
the necessary notation and 
introduce the two environments we use for our analysis, the weighted projective 
space $\textbf{WP}^4_{1,1,1,1,4}$ which we analyse near its conifold locus and 
the symmetric
torus $T^6$ both with one complex structure modulus. Both are chosen as simple 
toy examples and as they allow for a comparison with the performance of genetic 
algorithms on these environments~\cite{Cole:2019enn}.
\subsection{Flux datasets}
We consider a Calabi-Yau threefold $M$ with $h_{2,1}$ complex structure moduli 
and take s symplectic basis
$\{A^a,B_b\}$  for the $b_3 = 2(h_{2,1}+1)$ three-cycles, with 
$a,b=1,\dots,h_{2,1}.$\footnote{ This discussion 
mostly follows~\cite{Cole:2019enn} which is based on~\cite{DeWolfe_2005}. For a 
more detailed review of flux 
compactification see for example~\cite{Douglas_2007,Ibanez_2012}.} The 
cohomology elements $\alpha_a, 
\beta^b$ dual to our basis satisfy 
\begin{equation}
	\int_{A^a} \alpha_b = \delta_b^a\ , \qquad \int_{B_b} \beta^a = 
-\delta^a_b\ , \qquad \int_M \alpha_a \wedge\beta^b=\delta_a^b\ .
\end{equation}
The holomorphic three form $\Omega$, unique to our Calabi-Yau threefold, 
defines 
the periods\\ \noindent $z^a=\int_{A^a} 
\Omega,$ $ \mathcal{G}_b = \int_{B_b}\Omega$ and leads to the $b_3$-vector 
$\Pi(z)=(\mathcal{G}_b,z^a)$. It 
can also be written as $\Omega=z^a\alpha_a-\mathcal{G}_a\beta^a$. The Kähler 
potential for the complex 
structure moduli and the axio-dilaton $\phi = C_0 + ie^{-\varphi}$ is 
\begin{equation}
	\mathcal{K} = 
-\text{log}\left(i\int_M\Omega\wedge\bar{\Omega}\right)-\text{log}
\left(-i(\phi-\bar{\phi})\right)=-\text{log}
\left(-i~\Pi^\dagger\cdot\Sigma\cdot\Pi\right)-\text{log}\left(-i(\phi-\bar{\phi}
)\right)\ ,
\end{equation}
since $$\int_M\Omega\wedge\bar{\Omega} = 
\bar{z}^a\mathcal{G}_a-z^a\bar{\mathcal{G}}_a = 
-\Pi^\dagger\cdot\Sigma\cdot\Pi\ ,$$ with the symplectic matrix 
\begin{equation}
	\Sigma = \left(\begin{array}{c c} 0&{\bf 1}\\-{\bf 
1}&0\end{array}\right)\ .
\end{equation}
Turning on RR and NSNS 3-form fluxes leads to the quantized flux vectors in the 
$\alpha,\beta$ basis as
\begin{equation}
	F_3 = -(2\pi)^2\alpha'(f_a\alpha_a+f_{a+h_{2,1}+1}\beta^a),\quad H_3= 
-(2\pi)^2\alpha'(h_a\alpha_a+h_{a+h_{2,1}+1}\beta^a)\ ,
\end{equation}
where $f$ and $h$ are integer-valued $b_3$-vectors. In the following we set 
$(2\pi)^2\alpha'=1$ . The background 
fluxes introduce a superpotential~\cite{Gukov:1999ya}
\begin{equation}
	W = \int_M (F_3-\phi H_3)\wedge\Omega(z) = (f-\phi 
h)\cdot\Pi(z)\label{eq:sup}
\end{equation}
and a scalar potential
\begin{equation}
	V = e^\mathcal{K}(\mathcal{K}^{i\bar{\jmath}}D_iW 
D_{\bar{\jmath}}\bar{W}-3 
	|W|^2) \ .
\end{equation}
The $\mathcal{K}^{i\bar{\jmath}}$ is the inverse of the Kähler metric and 
$D_iW=(\partial_{i}+(\partial_{i}
\mathcal{K}))W.$ Here, the indices $i,\bar{\jmath}$ run over the complex 
structure moduli, the dilaton and the 
Kähler moduli. Due to the leading order no-scale structure for the Kähler moduli 
their contribution to the scalar potential cancels 
the $3|W|^2$ term and we end up with a positive semi-definite scalar potential 
\begin{equation}
	V = e^\mathcal{K}(\mathcal{K}^{a\bar{b}}D_aW D_{\bar{b}}\bar{W} +  
\mathcal{K}^{\phi\bar{\phi}} D_\phi W D_{\bar{\phi}}\bar{W})\ 
\label{eq:scalarpotential}.
\end{equation}
Here, the indices $a,\bar{b}$ only include the complex structure moduli and $\phi$ is 
the axio-dilaton. 
We are interested in vacua which have vanishing scalar potential (Minkowski 
vacua and fulfil the so called imaginary self-dual (ISD) condition). Hence we 
are looking for vanishing F-terms
\begin{eqnarray}
	D_\phi W &=&\frac{1}{\bar{\phi}-\phi}(f-\bar{\phi}h)\cdot\Pi(z)  = 0\ 
,\nonumber\\
	D_a W &=& (f-\phi 
h)\cdot(\partial_a\Pi(z)+\Pi(z)\partial_a\mathcal{K})=0\ .\label{eq:Fterm}
\end{eqnarray}
The fluxes contribute to the D3-brane charge via
\begin{equation}
	N_\text{flux} = \int_M F_3\wedge H_3 = f\cdot\Sigma\cdot h\ .
\end{equation}
For ISD fluxes we have $N_\text{flux}>0$. Since the total D3-brane charge on a 
compact manifold has to vanish 
and to ensure tadpole cancellation, appropriate negative charges have to be 
added (e.g.~by orientifolding). 
Effectively this introduces an upper bound  on $N_\text{flux}$~\cite{Cole:2019enn,DeWolfe_2005}:
\begin{equation}
	0<N_\text{flux}<L_\text{max} \label{eq:tadpole condition}\ ,
\end{equation}
where $L_{\rm max}$ is model dependent and we will utilise the same values as in 
\cite{Cole:2019enn} to keep 
comparability.
We combine our fluxes into the flux vector
\begin{equation}
	N = (f_1,\dots,f_{2h_{2,1}+2},h_1,\dots,h_{2h_{2,1}+2})^T\ . 
\label{eq:fluxvector}
\end{equation}

In summary, we search for flux vectors with $0<N_\text{flux}<L_{\rm max}$ and 
which yield vacua with vanishing scalar 
potential and additional phenomenological constraints on the expectation value 
for the dilaton and the superpotential. 
For this calculation, we now specify background geometries where we can 
calculate the period vector $\Pi(z)$ which 
allows us to calculate and solve the F-term conditions for the dilaton and 
complex structure 
moduli~\eqref{eq:Fterm}. 
Given them, we can go on to calculate explicitly the expectation values for the 
superpotential and the 
string coupling. 
In addition, the scalar potential is invariant under an $SL(2,\mathbb{Z})$ 
symmetry which acts non-trivially on the dilaton and the flux vectors. To avoid the resulting 
over-counting we restrict ourselves to 
dilaton values which lie in the fundamental domain that is
\begin{equation}
	\{\phi\} := \{\phi: |\text{Re}(\phi)|<0.5,\ |\phi|>1 \}\ . 
\label{eq:fdomain}
\end{equation}
\subsubsection*{Explicit background: Conifold in $\textbf{WP}^4_{1,1,1,1,4}$}
For the first experiments, we will work on a conifold described as a 
hypersurface in the weighted projective space 
$\textbf{WP}^4_{1,1,1,1,4}$ 
defined by
\begin{equation}
	\sum_{i=1}^4 x_i^8+4x_0^2+8\psi x_0x_1x_2x_3x_4 = 0\ 
\label{eq:conifold}.
\end{equation}
The Hodge numbers here are given by $h_{1,1}=1$ and $h_{2,1}=149$. We will 
consider the case of the orientifold 
$x_0\rightarrow -x_0$,~$\psi \rightarrow-\psi$ with worldsheet parity reversal 
that arises from F-theory compactified 
on a Calabi-Yau fourfold defined as a hypersurface 
$X_A=\textbf{WP}^5_{1,1,1,1,8,12}$ . Because of that special 
property, the tadpole can be calculated from the Euler characteristic of the 
fourfold to be \cite{Giryavets_2003}
\begin{equation}
	L_\text{max}=\frac{\chi(X_A)}{24}=\frac{23328}{24}=972 \ ,
\end{equation}
which sets the maximal tadpole for our flux configurations. This conifold 
\eqref{eq:conifold} has a symmetry group $\Gamma = 
\mathbb{Z}^2_8\times\mathbb{Z}_2$ under which 
all complex structure deformations are charged except $\psi$. Hence if we only 
turn on fluxes consistent with 
$\Gamma$, 
these charged moduli can be dropped and the periods can be calculated only for 
the axio-dilaton $\phi$ 
and the uncharged modulus $\psi$ \cite{Giryavets_2003}. Since we only have one 
complex structure modulus, the 
flux vector~\eqref{eq:fluxvector} is 8-dimensional and given by $N = 
(f_1,f_2,f_3,f_4,h_1,h_2,h_3,h_4)^T$. Near the 
conifold point $\psi=1$ the periods are \cite{Giryavets_2004}
\begin{eqnarray}
	\mathcal{G}_1(x) &=& (2\pi i)^3(a_0+a_1x+\mathcal{O}(x^2))\ , 
\nonumber\\
	\mathcal{G}_2(x) &=& \frac{z^2(x)}{2\pi i}\text{ln}(x)+(2\pi 
i)^3(b_0+b_1x+\mathcal{O}(x^2))\ ,\nonumber\\
	z^1(x) &=& (2\pi i)^3(c_0+c_1x+\mathcal{O}(x^2))\ ,\nonumber \\
	z^2(x)&=&(2\pi i)^3(d_0+d_1x+\mathcal{O}(x^2))\ ,\nonumber
\end{eqnarray} 
with $x=1-\psi,|x|\ll1$ \cite{Giryavets_2004}. The constants are given as 
\begin{eqnarray}
	a_0 &=& -1.774i, \quad a_1 = 1.227\ ,\nonumber\\
	b_0 &=& -1.047, \quad b_1 = 0.451 + 0.900i\ ,\nonumber\\
	c_0 &=& 4.952 - 5.321i, \quad c_1 = -4.488 + 3.682i\ ,\nonumber\\
	d_0 &=& 0.000, \quad d_1=1.800i\ .
\end{eqnarray}
Solving the F-term conditions \eqref{eq:Fterm} for the dilaton and complex 
structure modulus leads to \cite{Giryavets_2004}
\begin{eqnarray}
	\phi  
&=&\frac{f_1\bar{a}_0+f_2\bar{b}_0+f_3\bar{c}_0}{h_1\bar{a}_0+h_2\bar{b}
_0+h_3\bar{c}_0}+\mathcal{O}(|x|\text{ln}(|x|))\ ,\nonumber\\
	\text{ln}(x) & =& -\frac{2\pi i}{d_1}\Biggl(\frac{(f_1-\phi 
h_1)(a_1-\frac{\mu_1}{\mu_0}a_0)+(f_2-\phi 
h_2)(b_1-\frac{\mu_1}{\mu_0}b_0)}{f_2+\phi h_2}\nonumber\\
	&&\qquad\qquad+\frac{(f_3-\phi 
h_3)(c_1-\frac{\mu_1}{\mu_0}c_0)+(f_4-\phi h_4)d_1}{f_2-\phi h_2}\Biggr)-1\ . 
\label{eq:dilaton}
\end{eqnarray}
with 
\begin{equation}
	\mu_0 = 
i(2\pi)^6(a_0\bar{c}_0-c_0\bar{a}_0),\quad\mu_1=i(2\pi)^6(\bar{c}_0a_1-c_1\bar{a
}_0-d_1\bar{b}_0)\ .
\end{equation}
In this conifold background we are interested in vacua with a fixed absolute 
value for the expectation 
value of the flux superpotential $|W_0|=50,000.$ The motivation behind choosing 
such an artificial 
value is to estimate the ability to find vacua with nearby values. Such a tuning 
would for instance be 
required in tuning the cosmological constant in a LARGE volume 
scenario~\cite{Balasubramanian_2005}. 

In addition, for this value of the flux superpotential solutions are easily 
found even by random algorithms 
and it corresponds to the value the genetic algorithms in \cite{Cole:2019enn} 
have focused on.

\subsubsection*{Background geometry: symmetric torus}
\label{section:String_torus}
The second background in our experiments is the symmetric $T^6$ torus. We are 
also interested in 
specific values of the superpotential $W_0$ and in specific values for the 
string coupling $g_s.$ We  
follow the conventions of \cite{DeWolfe_2005}. \\
Since the considered torus is symmetric, it can be viewed as a direct product of 
three copies of $T^2$ 
and here we only have one complex structure modulus. Taking the axio-dilaton 
into account we get two 
moduli and in total 8 independent flux parameters. First, consider a non 
symmetric general $T^6$. The 
coordinates $x^i,y^i$ for $ i=1,2,3$ with periodicity $x^i\sim x^i+1,~y^i 
\sim y^i+1$ are defined such 
that the holomorphic 1-forms can be written as $d z^i = d x^i + \tau^{ij}d y^j$ 
with the complex 
structure moduli $\tau^{ij}$. The orientation is 
$$\int\ d x^1\wedge d x^2\wedge d x^3\wedge d y^1\wedge d y^2\wedge d y^3 = 1$$
and the symplectic basis for $H^3(T^6,\mathbb{Z})$ is 
\begin{eqnarray}
	\alpha^0 &=& d x^1\wedge d x^2\wedge d x^3\ , \qquad \alpha_{ij} = 
	\frac{1}{2} \epsilon_{ilm}d x^l\wedge d x^m\wedge d y^j\ ,\nonumber\\
	\beta^{ij}&=&-\frac{1}{2} \epsilon_{jlm} d y^l\wedge d y^m\wedge d 
	x^i\ , \qquad \beta^0=d y^1\wedge d y^2\wedge d y^3\ .
\end{eqnarray}
The holomorphic 3-form is given by 
$$\Omega=d z^1\wedge d z^2\wedge d z^3\ .$$
With this the 3-form fluxes can be expanded in terms of the symplectic basis 
\begin{eqnarray}
	F_3 &=& a^0\alpha^0+a^{ij}\alpha_{ij}+b_{ij}\beta^{ij} + b_0\beta^0\ , 
	\nonumber\\
	H_3 &=& c^0\alpha^0+c^{ij}\alpha_{ij}+d_{ij}\beta^{ij}+d_0\beta^0\ .
\end{eqnarray}
Now if we focus on the symmetric $T^6$ the complex structure moduli are all equal 
and we get 
$$\tau^{ij}=\tau\delta^{ij}\ .$$
In this case we now only have two moduli in total and the number of fluxes gets reduced as 
well and we have
\begin{equation}
	a^{ij}=a\delta^{ij}~,\quad b_{ij}=b\delta_{ij}~,\quad 
	c^{ij}=c\delta^{ij}~,\quad 
d_{ij}=d\delta_{ij}\ .
\end{equation}
The superpotential then only depends on the two moduli and takes the form
\begin{equation}
	W = P_1(\tau)-\phi P_2(\tau)
\end{equation}
with the polynomials 
\begin{eqnarray}
	P_1(\tau)&=&a^0\tau^3-3a\tau^2-3b\tau-b_0\ ,\nonumber\\
	P_2(\tau) &=& c^0\tau^3-3c\tau^2-3d\tau-d_0\ .
\end{eqnarray}
The Kähler potential simply is 
\begin{equation}
	\mathcal{K} = -3\log(-i(\tau-\bar{\tau}))-\log(-i(\phi-\bar{\phi}))
\end{equation}
and with this the F-term constraints are
\begin{eqnarray}
	P_1(\tau)-\bar{\phi}P_2(\tau)&=&0\ \label{eq:torus_fterm1},\\
	P_1(\tau)-\phi P_2(\tau)&=&(\tau-\bar{\tau})(P_1'(\tau)-\phi 
	P_2'(\tau))\label{eq:torus_fterm2}\ .
\end{eqnarray}
Solving Equation~\eqref{eq:torus_fterm1} for the dilaton leads to 
\begin{equation}
	\phi=\frac{\overline{P_1(\tau)}}{\overline{P_2(\tau)}}\ .
\end{equation}
The complex structure modulus $\tau=x+iy$ is then obtained by plugging this 
into 
\eqref{eq:torus_fterm2}. This leads to the equations
\begin{eqnarray}
	q_1(x)y^2&=&q_3(x)\ ,\\
	q_0(x)y^4&=&q_4(x)\ .
\end{eqnarray}
Eliminating $y$ and solving for $x$ leads to the cubic equation
\begin{equation}
	\alpha_3x^3+\alpha_2x^2+\alpha_1x+\alpha_0 = 0\ . \label{eq:cubic_torus}
\end{equation}
The exact form of the polynomials $q_i$ and the coefficients $ \alpha_i$ can be 
found in~\cite{DeWolfe_2005}. Since this equation is cubic, there will be up to 
three solutions for the 
moduli and hence up to three solutions for the superpotential as well. They are 
all equally valid and the 
algorithm will later on check all three of them. The D3-brane charge induced by 
the fluxes is
\begin{equation}
	N_{flux}=b_0c^0-a^0d_0+3(bc-ad)\ .
\end{equation}
As on the conifold we have to look for solutions where the dilaton lies in the 
fundamental domain 
(cf.~Equation~\eqref{eq:fdomain}). For the tadpole cancellation, one common 
orientifold for the torus is 
$T^6/\mathbb{Z}_2$ which has 64 O3-planes and hence $L_\text{max}=16$ 
\cite{Denef_2004}. We 
choose to work with $0<N_\text{flux}<16$ here. For the superpotential, we will 
look for values $|W_0|<10$ and we will also perform experiments to search 
for $g_s\approx 0.3$.

\subsection{RL environment and reward structure}
We are interested in exploring the flux vacua in these two background 
geometries. To explore them with 
RL we have to specify how we can navigate between various flux vacua. 

We implement these environments with the use of OpenAI gym~\cite{openaigym}. For 
that purpose we 
overwrite the following three functions:
\begin{itemize}
	\item \textbf{step}: The function to actually move through the 
environment. It takes a specific action 
	and a state as an input and uses it to transition to 
	a new state. This state is then 
	returned together with the reward for that transition and an indication 
whether the episode is over 
	and if all conditions were satisfied.
	\item \textbf{reset}: The function to reset the environment to its 
initial configuration. This will be 
	called at the beginning of each episode.
	\item \textbf{seed}: The function to seed the random number generators. 
Having a seed will give 
	the same sequence of random numbers for the same initial data. 
\end{itemize}
Gym comes with two possible spaces, a continuous {\it Box space} and a discrete 
{\it Discrete space}. 
We will define the state space as an 8 dimensional {\it Discrete space}, which 
corresponds to our flux 
vector in equation \eqref{eq:fluxvector}. The action space will be a 16 
dimensional {\it Discrete space}, 
since we allow the agent to raise or lower one of the flux quanta in the flux 
vector by $\pm1$. So for 
example action $a=2$ will raise the second entry of the flux vector by one and 
action $a=10$ will lower 
that entry by one. The last thing to define are the reward functions. 
\subsubsection*{Reward functions}
Throughout the experiments the reward functions were changed to improve 
performance. Here we 
present the general idea and describe later our hyperparameter tuning. As 
mentioned earlier, three 
conditions are checked for all of which we introduce a reward:
\begin{enumerate}
	\item \textbf{Gauge condition}: $\phi\in\{\phi: |\text{Re}(\phi)|<0.5,\ 
	|\phi|>1 \}$ dilaton in 
	fundamental domain.
	\item \textbf{Tadpole condition:} whether $0<N_\text{flux}<972$ on the 
	conifold or $0<N_\text{flux}<16$ on the torus.
	\item \textbf{Superpotential condition:} $|W_0|=50000\pm1000$ on the 
	conifold and $|W_0|<10$ on the torus.
\end{enumerate}
For fulfilling the gauge and the superpotential conditions, a fixed reward 
$r_\text{gauge}$ and 
$r_\text{sup}$ respectively is given. For not fulfilling the tadpole condition, 
a negative reward 
(punishment) $-r_\text{tad}$ is given. Hence the agent receives a maximal reward 
of $r_\text{gauge} + 
r_\text{sup}$. Since all of these conditions are hard to satisfy, the agent will 
receive no feedback 
whatsoever during his exploration of the environment. To avoid this and hence 
speed up learning, 
rewards (or punishments) were given proportional to the distance to the optimal 
solution. In more detail 
they look like
\begin{eqnarray}
	r_g&=& 
	\begin{cases}
		N_g\cdot 
		
\text{exp}\left(-\left(\frac{|\text{Re}(\phi)|-0.5}{\sigma_g}
\right)^2\right)\cdot
		 r_{\rm gauge} \ &\text{if Im}(\phi)>1\ ,\\
		N_g\cdot 
		
\text{exp}\left(-\left(\frac{|\text{Re}(\phi)|-0.5}{\sigma_g}
\right)^2\right)\cdot

\text{exp}\left(-\left(\frac{\text{Im}(\phi)-1}{\sigma_g}\right)^2\right)\cdot
		 r_{\rm gauge}\ &\text{else}\ ,
	\end{cases}\\
	r_{t} &=& N_t\cdot\text{tanh}\left(\sigma_t\left( 
	N_\text{flux}-0\right)\right)\cdot r_\text{tad}\ ,\\
	r_s &=& 	
N_s\cdot\text{exp}\left(-\left(\frac{|W_0|-50000}{\sigma_s}\right)^2\right)\
	 ,\label{eq:rs}
\end{eqnarray}
with normalization constants $N_i$ and standard deviations $\sigma_i$. The latter 
control how far away 
from the optimal solution a state can be and still be rewarded. There are a few 
things to note here. First, 
it is hard to define a distance from the fundamental domain. Therefore, we just 
choose this product of 
individual distances. Second the tadpole punishment is given as a tanh function, 
since here the 
punishment should get smaller the closer one gets to the optimal solution. The 
distance is only 
calculated to 0 because states with a tadpole larger then 972 are basically not 
present. Third, in the 
superpotential distance reward $r_s$ the exponential is not multiplied with the 
superpotential reward 
$r_\text{sup}$. If a correct superpotential value is found, the episode is 
ended. Hence, a state with a 
correct superpotential value is a terminal state. If the difference in reward 
between the terminal state 
and the state before is too small, it would be more rewarding for the agent to 
run around that terminal 
state forever never hitting it and hence never finding an optimal solution. To 
ensure that this reward 
difference is big enough, the exponential in $r_s$ is not multiplied with 
$r_\text{sup}$.
\begin{figure}
	\includegraphics[width=\textwidth]{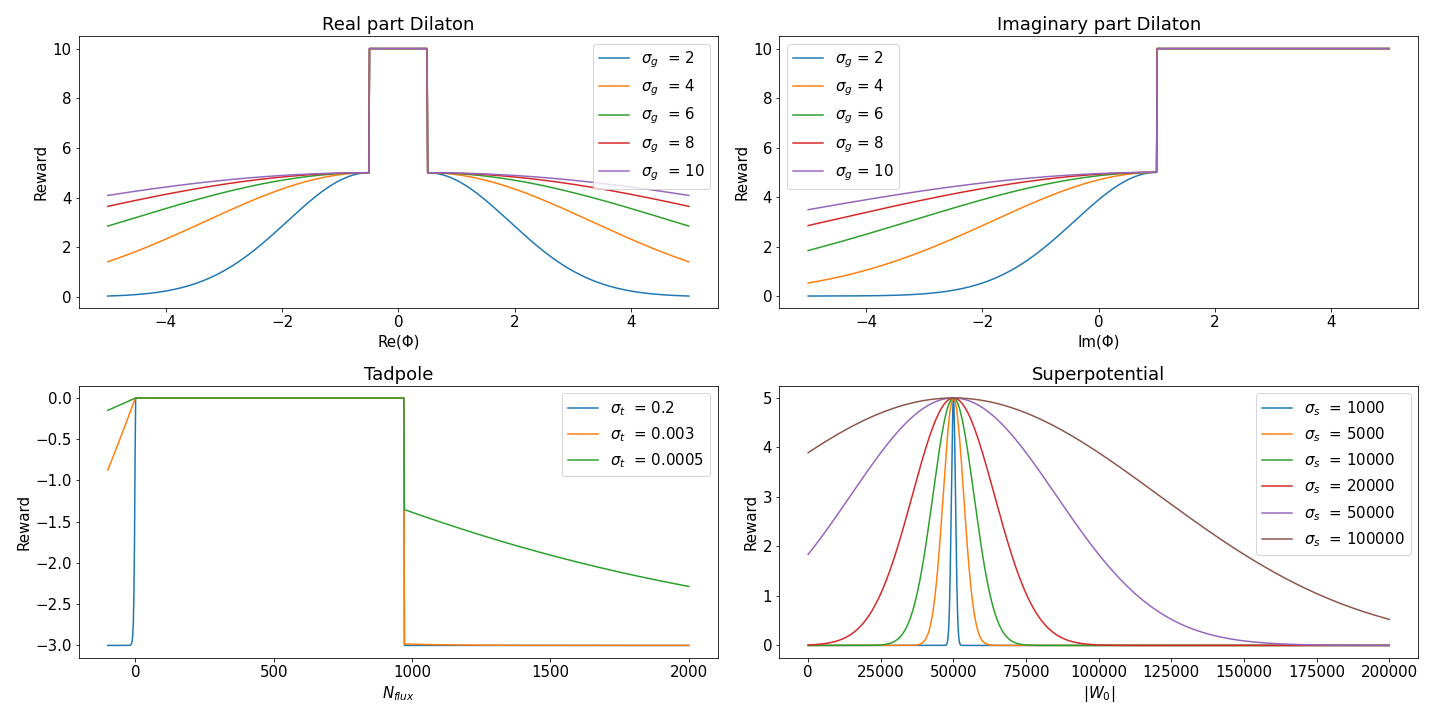}
	\caption{Reward function 
	for different $\sigma$ values. 
	The first two plots show the reward functions for the real and imaginary 
part of the dilaton, while 
	the third one shows the tadpole and the last shows the superpotential. 
The $\sigma$'s have to be 
	chosen to balance between distinguishing states near the maximum and 
still covering a large area 
	of possible values for the physical quantities. For the superpotential 
the 
	agent receives a fixed reward of 7000 if $49000<|W_0|<51000$. }
	\label{fig:rewards}
\end{figure}
The reward functions for different $\sigma$ values are shown in 
Figure~\ref{fig:rewards}. When 
choosing the $\sigma$ value, we have to balance between a large enough value to 
make sure that a 
large part of the environment is rewarded and a small enough value to still 
distinguish different states 
near the optimal solution. 
\subsubsection*{RL agents and neural networks}
We use an advantage actor-critic agent (A3C) \cite{mnih2016asynchronous} and 
double deep Q-learning with 
prioritized experience replay and duelling extensions 
\cite{Mnih_2015,Wang_2016,schaul2016prioritized,vanhasselt2015deep}. A short 
introduction to 
these 
algorithms can be found in Appendix~\ref{app:RL}.\footnote{For the A3C 
implementation we use ChainerRL \cite{chainer_learningsys2015} and the DQN algorithm is implemented in 
PyTorch \cite{paszke2019pytorch}. } 
For more details about 
reinforcement learning see~\cite{Sutton1998}.
We use a dense network for all of our experiments, which is inspired 
by~\cite{Halverson_2019}. It 
consists of four hidden layers, each with ReLu activation. For the A3C setup the 
activation function for 
the output layer of the policy network will be a softmax function to get 
probabilities for each action. The 
output will be a vector with dimension corresponding to the number of actions so 
each entry 
corresponds to one specific action. The first three layers have size 50 while 
the last layer has size 200. 
\section{Experiments/Explorations}
\label{sec:experiments}
Our RL experiments have been designed with the following scope. We are 
interested whether our RL 
agents are able to learn what successful vacua are and whether they can reveal 
structures on the 
space of vacua. Such structures will be in correlations among the flux quanta 
and how the RL agent 
navigates through the flux environment starting from a random starting point. As 
these RL methods are 
computationally more demanding than standard random or metropolis algorithms, we 
compare the efficiency of these 
three approaches. For this comparison we present our best working agents on the 
conifold 
environment. We present the results of our experiments for the metropolis 
algorithm, random walker, 
A3C and DDQN with prioritized experience replay and duelling extensions.

Our training proceeds as follows. A model is found, if all of the three 
conditions mentioned earlier in 
section \ref{sec:fluxenvironments} (gauge, tadpole, superpotential) are 
satisfied. The agents are run for 
a given number of steps $N_\text{steps}$ and reset if this step number is 
reached or a model is found. 
A step corresponds to one call of the step function in the environment that 
means a transition from one 
state to another e.g$.$ going from $N = [1,0,0,0,0,0,0,0]$ to $N = 
[2,0,0,0,0,0,0,0]$. This will be 
repeated for a number of episodes $N_\text{episode}$. Successful models can be 
distinguished by their 
final flux vector as we are restricting ourselves to vacua in the fundamental 
domain. All fluxes are 
initialized in the interval $[-30,30]$ where the starting point is drawn from a 
uniform distribution.

To specify the A3C we have to choose certain hyperparameters. We used for the 
learning rate $l_r=5\cdot10^{-4}$, for the discount factor $\gamma=0.95$ and 
for the exploration parameter $\beta=0.1$. 

For prioritized duelling DDQN we use
the learning rate $l_r=5\cdot10^{-6}$ and discount factor $\gamma=0.999$. The 
other relevant parameters are $\alpha = 0.1\ , \beta=0.6$ for 
prioritized experience replay and $\epsilon_\text{end} = 0.01\ 
,\epsilon_\text{decay} = 0.999$ for the epsilon greedy strategy. More details 
can be found in Appendix~\ref{app:RL}.

Our RL results will always be compared to a random walker and a metropolis 
algorithm which 
both were run 
under the same circumstances as the plotted agents meaning with the same number 
of steps and the 
same reward structures. The random walker just chooses a random action at each 
step. In case of the 
metropolis algorithm, the agent chooses an action at random as well, but only 
transitions to the next 
state to which this action would lead to, if the reward of that state is higher 
than the one for the state he 
currently is in. If the reward is lower, the transition only happens with a 
probability depending on the 
difference in rewards
\begin{equation}
 P = e^{-(r_{s+1}-r_s)}\ .
 \end{equation}
 
\begin{figure}
\begin{center}
	\includegraphics[width=\textwidth]{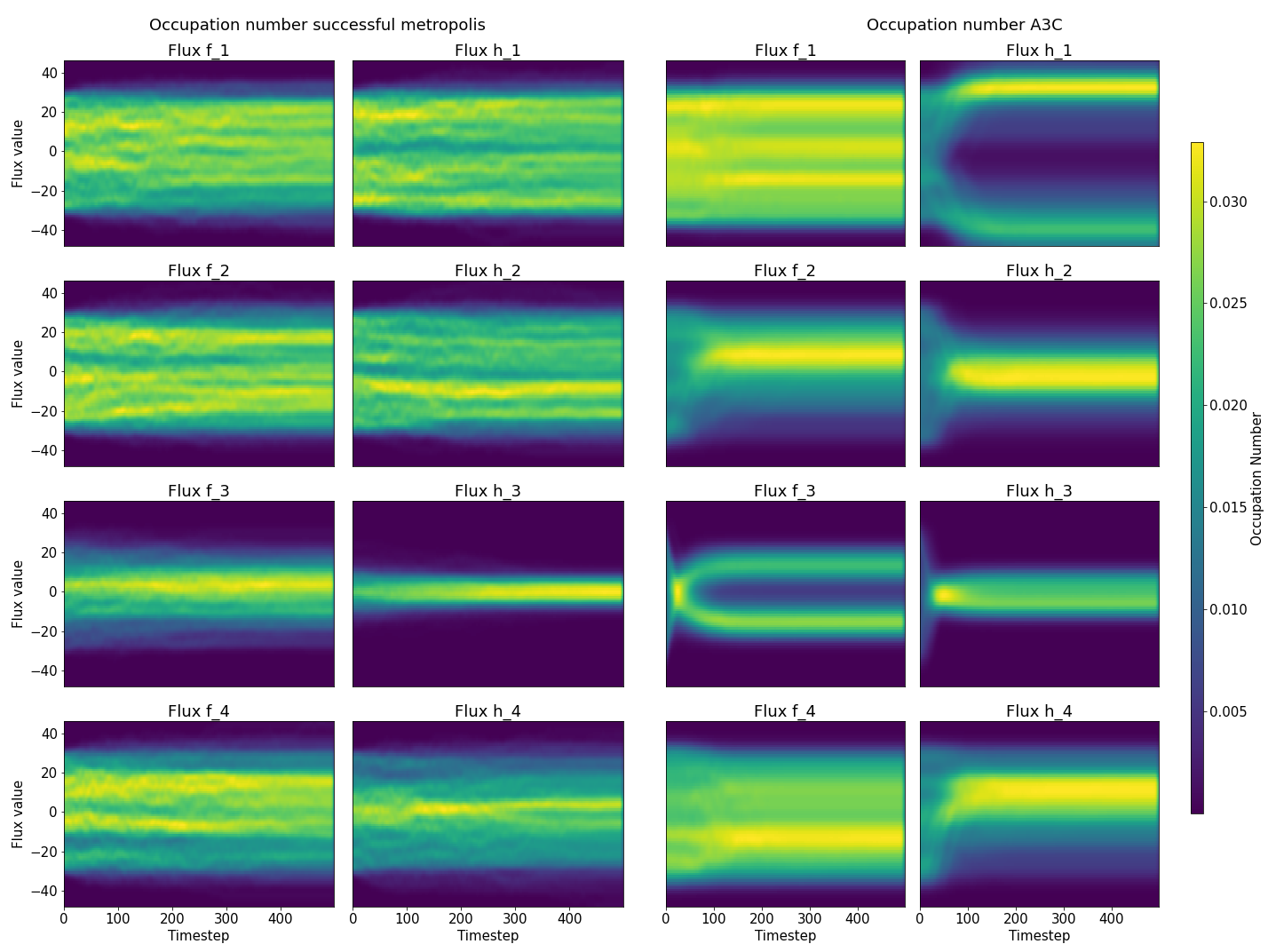}
	\end{center}
	\caption{Here the occupation number of each flux value is shown for the 
successful metropolis 
runs and the A3C. Every plot corresponds to a flux number. Then the colour 
represents how often, in 
500 runs, a given value of this flux number occurred at a given timestep. The 
flux values for the 
algorithms were initialized between -30 and 30.}
	\label{fig:met_occ}
\end{figure}

To establish a basis of comparison let us first describe the explorations with 
the random walker and a 
metropolis algorithm. In general they both behave completely random meaning that 
no structure in the 
flux values is visible. However, if we only focus on the successful runs meaning 
the ones where a good 
vacuum was found, the situation changes. In that case we show the occupation 
number in our runs in 
Figure~\ref{fig:met_occ}.
We depict the occupation number 
\begin{equation}
	O_{i,j}(t) = \frac{N_{i,j}(t)}{N_a(t)}\ ,
\end{equation}
where $N_a(t)$ is the total number of agents at timestep $t$ and $N_{i,j}(t)$ 
the number of agents 
which reach the flux value $i$ of flux number $f_j$ or $h_j$ at timestep $t$. 
Here $N_a(0) = 500$ hence 
500 runs where done. We find that $h_3$ is oriented more closely towards zero.
\begin{figure}
\begin{center}
	\includegraphics[width=\textwidth]{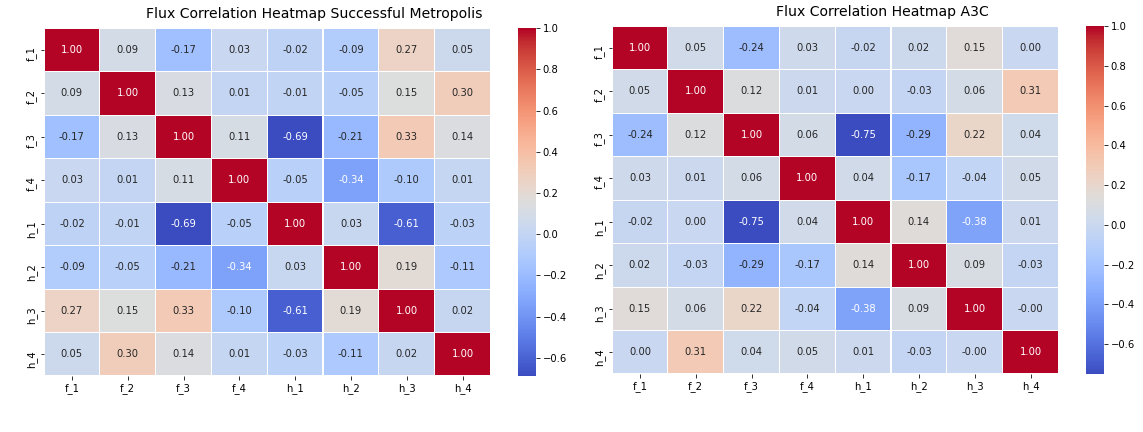}
	\end{center}
	\caption{This plot shows the correlation heatmap of the different flux 
numbers for the successful 
metropolis algorithm and the A3C. The value in each entry is just the 
correlation between all 
encountered flux values of this flux numbers. A positive value and red colour 
means that the two fluxes 
are correlated whereas blue and negative means they are anti-correlated.}
	\label{fig:random_heatmap}
\end{figure}
More structure can be seen in the plot from our A3C agents. The agents seems to 
not touch $f_1$ and 
$f_4$ at all, whereas $f_2$ is kept a little bit above zero, $h_2$ a little bit 
below zero, and $h_3$ 
around zero while $h_4$ is raised most of the times. With the correlation map in 
mind, we can safely 
say that $f_3$ and $h_1$ are pushed to opposite high or low values. $f_3$ and 
$h_1$ are pushed to 
opposite high or low values which we confirm below when we discuss the 
correlations among the flux 
quanta. In general, the heat map has similarities with the metropolis one where 
only the successful runs 
were plotted but features additional structures.

Correlations among the flux vacua of vacua can be seen in the correlation heat 
map shown in Figure~\ref{fig:random_heatmap}. Here we find a strong 
anti-correlation between $f_3$ and $h_1$ as the most dominant feature. These 
correlations in phenomenologically interesting vacua have not been pointed out 
before.

 Since these structures only appear for successful runs, we conclude that they 
are inherent environmental features. 
\begin{figure}
	\includegraphics[width=\textwidth]{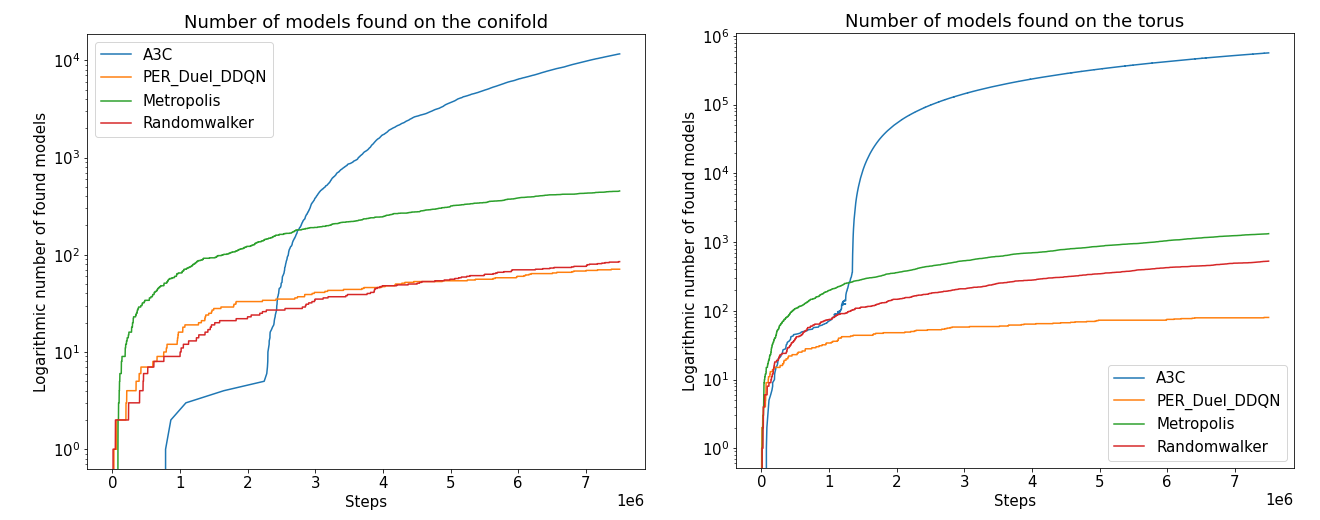}
	\caption{Comparison of log number of  distinct models found per 1000 
Episodes of A3C with PER dueling DDQN. A3C shows the best and fastest learning 
behaviour whereas PER Duel DDQN does not even beat the random walker. The flux 
values for all algorithms were initialized between -30 and 30.}
	\label{fig:perepisode_dist}
\end{figure}

Another measure of the success is given by the number of models found during 
training over the number of training steps which is shown in 
Figure~\ref{fig:perepisode_dist}. We clearly see that the A3C RL implementation 
finds $\mathcal{O}(20)$ more distinct models than metropolis and even order 
$\mathcal{O}(200)$ more than a random walker. Hence the agent is way more 
efficient in finding good vacua than metropolis. This is even more apparent when 
considering the number of steps needed to find $100$ models. On the conifold, 
the agent needs $\sim10^4$ steps while metropolis needs $\sim1.2\times10^6$ and 
a random walker $\sim7.7\times10^6$. So our agent is of order $\mathcal{O}(100)$ 
better then metropolis and order $\mathcal{O}(800)$ better than a random walker. 
The reason is, that he exploits the structures in the environment.

\begin{figure}[t]
	\includegraphics[width=\textwidth]{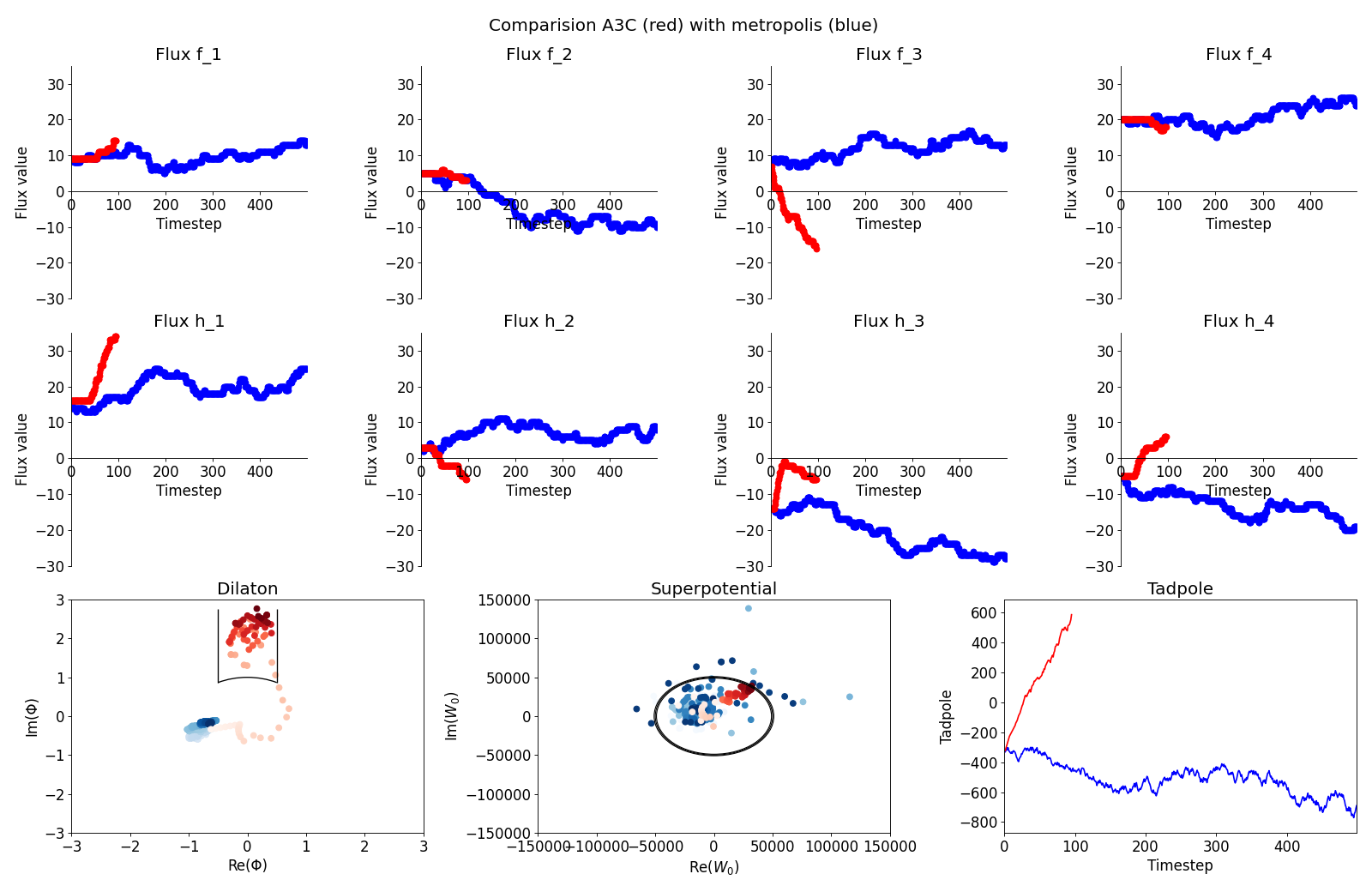}
	\caption{Here we see the result of one agent and a metropolis algorithm 
running through the environment. The A3C is shown in red and metropolis in blue. 
The plot shows the flux values they encounter as well as the dilaton and 
superpotential values. For the dilaton, the black line indicates the fundamental 
domain whereas for the superpotential the black circle defines the area of 
values we are looking for. The fluxes for the agent were initialized between -30 
and 30. Also compared to metropolis, the A3C agent manages to find a model 
rather quickly and to remain inside the fundamental domain.}
	\label{fig:met_a3c_comp}
\end{figure}

We can understand this by displaying the path from a given random configuration 
which are taken by the A3C agent and the metropolis approach respectively which 
is displayed in Figure~\ref{fig:met_a3c_comp}.
Both algorithms are initialized at the same flux vector and then run for up to 
500 steps or until they found a good vacuum. We clearly see, that our agent 
first sets $h_3$ close to zero and then pushes $h_1$ and $f_3$ to opposite 
directions. This is achieved in a very fast way and hence the A3C agent finds a 
model way earlier then metropolis who does not succeed at all in this example. 
By closely inspecting the movement of the dilaton value in correspondence to the 
flux movements, we see that setting $h_3$ to zero is mainly for transitioning 
into the fundamental domain. We also notice, that our agent manages to almost 
never step outside of the domain again, whereas metropolis just walks in and out 
randomly. Also, our agent keeps the superpotential close to 50000 and is 
stepping carefully around it until he hits the desired value. Clearly this is 
for a specific initialization but we have observed this in several instances and 
note that overall the A3C finds significantly more flux vacua (cf. 
Figure~\ref{fig:perepisode_dist}).

On the torus, our agent finds order $\mathcal{O}(300)$ more models during 
training than metropolis and order $\mathcal{O}(1000)$ more than a random 
walker. The same can be found if we look at the number of steps needed to find 
100 models. In this case, our trained agent needs $\sim10^3$ steps while 
metropolis needs $\sim3\cdot10^5$ steps and a random walker needs $\sim10^6$ 
steps. Hence A3C is again of order $\mathcal{O}(300)$ better than metropolis and 
of order $\mathcal{O}(1000)$ better than a random walker. We remark though, that 
these numbers do not hold for distinct models. The A3C only finds order 
$\mathcal{O}(2000)$ distinct models during training as does a metropolis 
algorithm. In \cite{Denef_2004} it was found, that the number of vacua is 1231, 
while in \cite{DeWolfe_2005} the number scales with $0.13~L_\text{max}$ for 
$L_\text{max}>20$ and shrinks even more rapidly than $L^4$ for $L<20$. Hence, we 
do not expect many more vauca than the identified order $\mathcal{O}(2000)$. So 
our trained agent does not find more distinct models than a metropolis algorithm 
because there are not many more to find, but he runs towards them in a way more 
efficient and faster way. The shown experiments on the torus were done on a flux 
initialization interval of $[-3,3]$, because for a larger one like $[-30,30]$ a 
random walker and even a metropolis algorithm do not find any good vacua while 
running for $7.5\cdot10^6$ steps. Therefore A3C's learning is substantially 
hindered and he also does not manage to find a single model. If however, the 
agent trained on the interval $[-3,3]$ is run on the interval $[-30,30]$ he does 
find good vacua. He recognises where in the environment he is and manages to run 
towards small flux values where he knows all the models are. Metropolis is not 
able to do anything similar.

\subsubsection*{Search for $g_s$}
\begin{figure}[t]
\begin{center}
	\includegraphics[width=0.49\textwidth]{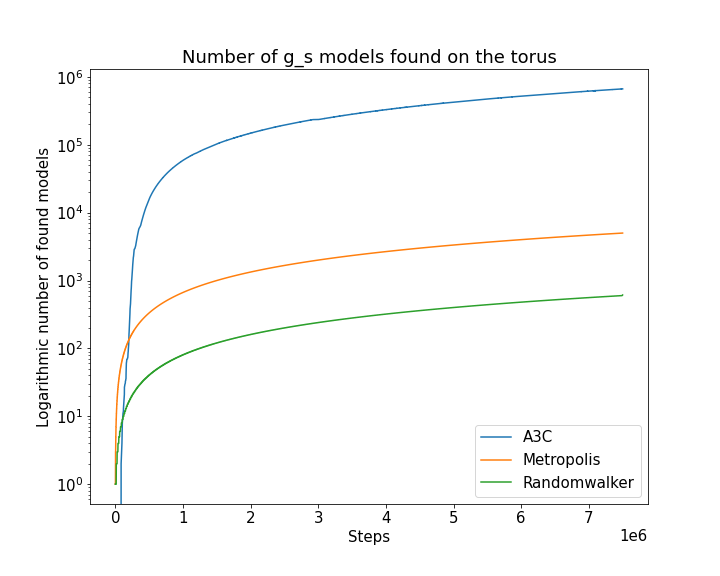}
	\end{center}
	\caption{Plot of log number of models of A3C on the 
		torus. He searches for string coupling values with $0.29<g_s<0.4$. A3C shows a 
		good 
		learning behaviour and outperforms the others strongly.}
	\label{fig:a3ctorus gs}
\end{figure}
On the torus we also look for solutions with $0.29<g_s<0.4$. Since $g_s = 
\frac{1}{\text{Im}(\phi)}$ this corresponds to looking for 
$2.5<\text{Im}(\phi)<4 $. Because we still do not want to over count and 
we want physically meaningful solutions, we will also imply the fundamental 
domain condition $|\text{Re}(\phi)|<0.5$ and the tadpole condition 
\eqref{eq:tadpole condition}. Again, our algorithm will calculate the dilaton for 
all real solutions of~\eqref{eq:cubic_torus}. The neural network and environment 
structures are the same as for the superpotential search as are the 
hyperparameters. The distance rewards (punishments) are given by
\begin{eqnarray}
	r_g&=& 
	\begin{cases}
		N_g\cdot 
		
\text{exp}\left(-\left(\frac{|\text{Re}(\phi)|-0.5}{\sigma_g}\right)^2\right)
		 \ &\text{if Im}(\phi)>1\ ,\\
		N_g\cdot 
		
\text{exp}\left(-\left(\frac{|\text{Re}(\phi)|-0.5}{\sigma_g}
\right)^2\right)\cdot

\text{exp}\left(-\left(\frac{\text{Im}(\phi)-3}{\sigma_g}\right)^2\right)
		 \ &\text{else}\ ,
	\end{cases}\\
	r_{t} &=& N_t\cdot\text{tanh}\left(\sigma_t\left( 
	N_\text{flux}-0\right)\right)\cdot r_\text{tad}\ ,
\end{eqnarray}
with the constants
\begin{eqnarray}
	N_g = 5\ , &\quad \sigma_g = 5\ , &\quad r_\text{gauge}= 2000 \ , \\
	N_t = 1\ , &\quad \sigma_t=0.2\ , &\quad r_\text{tad} = -5\ .
\end{eqnarray}
As RL agent, we used only the A3C since PER Duel DDQN did not succeed for the 
superpotential search. The agent is again compared to a random walker and a 
metropolis algorithm. The agent was trained for $5\cdot10^6$ steps and reset 
after he found a model or after 500 steps. The results are plotted in Figure 
\ref{fig:a3ctorus gs}. The A3C finds significantly more solutions than the 
other two algorithms. To find 100 models it takes our agent again $\sim10^3$ 
steps while metropolis needs $\sim10^5$ steps and a random walker $\sim10^6$. 
A3C is more successful because he again exploits efficiently that good vacua 
are to be found in a region of small flux values. More details about the torus experiments can be 
found in Appendix~\ref{app:torusenvironments}.\\
 
To sum up, the A3C agent learns and exploits very efficiently the apparent 
structure in these specific environments. We did not know about this structure 
before, so an agent like this could be used on different and more complex 
environments such that he could help us there to find structures we did not know 
about and help us to learn more about the string landscape.\\
We note that prioritized dueling DDQN only manages to beat the 
random walker on the conifold and never beats a metropolis algorithm. We have 
performed an extensive hyperparameter search (see 
Appendix~\ref{app:hyperparameters} for more details). 

\section{Conclusions}
\label{sec:conclusions}
Understanding the landscape of flux vacua is a long-standing problem in string 
phenomenology. Estimates on the number of vacua require an efficient sampling 
technique to make progress towards understanding the landscape.\footnote{This 
rich structure of solutions is not only intrinsic to flux compactifications but 
has also been noted significantly earlier in the context of heterotic string 
theory~\cite{Lerche:1986cx}.} We find that for few number of moduli standard 
metropolis explorations are already feasible to identify the correlations among 
different flux vacua.
We found that our RL agents are able to explore a phenomenologically distinct 
area of flux space by 
providing it with an appropriate reward structure. Generally speaking, after 
sufficient training, the RL algorithm can outperform the metropolis explorations 
significantly. The main question this raises are how these methods scale with 
the complexity of the underlying model and whether RL algorithms remain 
trainable and enable in those cases significant new insights which are not 
attainable via metropolis sampling.

In addition we show that we are able to explore different vacua in a small 
region 
of flux space. This is crucial as the associated correlations in flux space will 
hold valuable information on 
which structures are needed for addressing the cosmological constant problem via 
the string landscape.

There are multiple avenues how to extend this work:
\begin{itemize}
 \item We have restricted ourselves to vacua where analytic solutions where 
available. In most cases 
such analytic solutions are not available and one needs to search for such vacua 
numerically. In 
principle the two can be combined which will increase the runtime and 
potentially misses some vacua. 
Examples of such numerical searches can be found for instance 
in~\cite{MartinezPedrera:2012rs,Cicoli:2013cha}.
 \item Do the configurations of phenomenologically interesting flux vacua reveal 
an interesting 
sub-structure which are not visible in the flux-flux correlations? It will be 
interesting to examine the search for such structures using topological data 
analysis~\cite{Cirafici:2015pky,Cole:2018emh}.
 \item In terms of scaling of these RL techniques, one parameter which is of interest is the 
initial range of flux 
vacua which determines roughly speaking how many vacua are nearby and whether 
the algorithm 
learns to explore a large fraction of distinct vacua or always collapses to the 
same ones. Our work has 
focused on type~IIB string theory which captures only a subset of fluxes in 
comparison to the situation 
in F-theory where many more flux vacua are present where this scaling question 
becomes more 
pressing (cf.~\cite{Taylor:2015xtz} for an estimate of $10^{272000}$ flux vacua 
on a fixed background geometry and estimates on the number of background 
geometries~\cite{Halverson:2017ffz,Taylor:2017yqr}).
 \item At this moment we have to adapt our environment and reward structure for 
each background 
geometry. For a large exploration of the string landscape it seems crucial to 
being able to apply the 
same strategy across various background geometries which has the potential to 
reveal universal 
features in the identification of phenomenologically interesting vacua.
\item Another approach for revealing structures is by looking at generative techniques of effective field theories satisfying certain UV~\cite{Erbin:2018csv,Halverson:2020opj} or IR constraints~\cite{Hollingsworth:2021sii}. Again it would be very useful to understand which methods are best suited to identify systematics in phenomenologically interesting string vacua. In particular this is to identify the best suited methods for these sparse solutions in high-dimensional spaces.
\end{itemize}
We hope to return to these questions in the future.

\section*{Acknowledgments}
We would like to thank Alex Cole and Gary Shiu for useful discussions.

\appendix
\section{RL-details}
\label{app:RL}
Here we provide a lightning review of the RL algorithms which we are using. 
Deep Q learning and its extensions are based on 
\cite{Mnih_2015,Wang_2016,schaul2016prioritized,vanhasselt2015deep}.
\subsection{Deep Q-learning}
Here the state-action value function depends on the weights of a neural network 
\begin{equation}
	Q^\pi(s,a) \simeq Q^\pi_\theta(s,a)\ ,
\end{equation}
with a state $s$, an action $a$ and the network weights $\theta$. For Deep 
Q-learning (DQN) one initializes two different sets of weights $\theta$ and 
$\theta'$ and only updates the so called target weights $\theta'$ after a given 
number of timesteps. The update rule for $\theta$ then looks like
$$ \Delta\theta=\alpha(r_{t+1}+\gamma\cdot 
\text{max}_aQ^\pi_{\theta'}(s_{t+1},a)-Q^\pi_\theta(s,a))\nabla_\theta 
Q^\pi_\theta(s,a)$$
with learning rate $\alpha$, reward $r_{t+1}$, discount factor $y$ and $\theta' 
= \theta$ is set after 
a given number of timesteps. To maneuver 
through the environment, the agent then follows a policy
\begin{equation}
	\pi(s) = \text{argmax}_a Q_\theta(s,a) \ ,
\end{equation}
which is called a greedy policy. Following this policy is equivalent to purely 
exploiting. 
To get some exploration as well, one usually uses a so called $\epsilon$-greedy 
policy. Here the agent only chooses the best action with probability 
$1-\epsilon$ and with probability $\epsilon$ a random action. Usually one wants 
to explore more at the beginning of the training and chooses a start value of 
$\epsilon_{start}=1$ and exploit more at the end so one decreases epsilon at 
every time step by multiplying it with a small number like 
$\epsilon_{decay}=0.99985$ until a final (small) value for example 
$\epsilon_\text{end} = 0.01$. The last piece we need to arrive at the commonly 
used DQN algorithm is experience replay. By moving through the environment the 
agent sees different tuples of $(s,a,r,s')$ of state, action, reward and next 
state. This will lead to a trajectory in the environment represented by a 
series of tuples$$\tau = ((s_0,a_0,r_0,s_1),(s_1,a_1,r_1,s_2),\dots,( 
s_{T-1},a_{T-1},r_{T-1},s_T))\ .$$ The latest trajectories (latest $C$ tuples) 
will be stored in a so called replay buffer $D$. To then update the Q-function, 
a number $N$ of these tuples is sampled uniformly at random from the replay 
buffer. This avoids data inefficiency and correlation between the tuples inside 
one trajectory.
\begin{algorithm}[h]
	\begin{algorithmic}[1]
		\State Initialize replay buffer $D$ with a fixed capacity C
		\State Initialize state-action value function $Q$ with random 
weights 
		$\theta$
		\State Initialize target state-action value function $Q$ with 
weights 
		$\theta'=\theta$
		\State Initialize $\epsilon$ to start value 
$\epsilon=\epsilon_{start}$ 
		\For{episode $m=1,\cdots,M$}
		\State Observe initial state $s_1$
		\For{timestep $t=1,\cdots,T$}
		\State Select action $a_t = \begin{cases}  \text{random action 
with 
		probability } \epsilon \\ \text{argmax}_a Q_\theta(s_t,a) 
\text{ 
		otherwise} \end{cases}$
		\State Execute action $a_t$ in environment, observe reward 
$r_{t}$ and 
		next state $s_{t+1}$ and store transition $(s_t, a_t, r_t, 
s_{t+1})$ in 
		$D$
		\State Sample uniformly a random minibatch of $N$ transitions  
$(s_j, 
		a_j, r_j, s_{j+1})$ from $D$
		\State Set $y_j= \begin{cases} r_j \text{ if episode ends at 
step } 
		j+1\\ r_j + \gamma \text{max}_a Q_{\theta'}(s_{j+1},a) \text{ 
		otherwise} \end{cases}$
		\State Update parameters $\theta$ according to $\Delta\theta 
		=\frac{1}{N}\sum_{j=1}^N \alpha (y_j - 
Q_\theta(s_j,a_j))\nabla_\theta 
		Q_\theta(s_j,a_j)$
		\State Set $\epsilon \leftarrow \text{min}(\epsilon_{end}, 
		\quad\epsilon\cdot\epsilon_{decay}) $
		\State Every $X$ steps set $\theta'=\theta$
		\EndFor
		\EndFor
	\end{algorithmic}
	\caption{DQN algorithm}\label{alg:DQN}
\end{algorithm}
The whole algorithm is given in Algorithm \ref{alg:DQN}.
\subsection*{DQN improvements} 
There are several performance improvements to DQN which will be covered in the 
following. 
\subsubsection*{Double Deep Q-learning (DDQN)}
DQN works in general, but suffers from the problem of overestimating values. 
The reason is that the target in Line 11 in Algorithm \ref{alg:DQN} is given by 
the maximum of the same Q function as the action is chosen from in line 8. This 
maximum operator comes from the idea, that the best action in the next state is 
given by the one with the highest Q value. But that is not necessarily the case 
and just an estimation. Since this estimation is done twice, once in line 8 and 
once in line 11, overestimation is present and the estimated values will be 
overoptimistic. To overcome this problem, one uses two different networks which 
in our case are the ones with weights $\theta$ and $\theta'$. Then the target  
in line 11 for DDQN becomes $$y^{DDQN}_j = r_j + \gamma 
Q_{\theta'}\left(s_{j+1},\text{argmax}_aQ_\theta(s_{j+1},a)\right)\ .$$ The 
rest of the algorithm is the same as for DQN. There are further extensions to 
DDQN two of which will be discussed in the following.
\subsubsection*{Dueling} 
Dueling DQN differs from normal DQN in the network architecture. Here, before 
outputting the Q-function, the network will be split in two, the so called 
advantage function $A(s,a) = Q(s,a)-V(s)$ and the value function $V(s)$. The 
motivation for this is that it is not always necessary to know the value of 
each action choice but just the state value instead. The Q function is then 
again obtained by adding up the split layers and subtract the average advantage
\begin{equation}
	Q(s,a) = V(s) + A(s,a) - \frac{1}{\mathcal{A}}\sum_{a'} A(s,a')\ ,
\end{equation}
with $\mathcal{A}$ the dimensionality of the action space. The subtraction is 
needed, since otherwise $V$ and $A$ can not be recovered uniquely from the 
Q-function, leading to problems in backpropagation.
\subsubsection*{Prioritized experience replay} 
At every training step a set of states the agent has visited during training is 
sampled and given to the network in order to calculate and update the Q-values 
for these states. The idea of prioritized experience replay is to assign a 
probability to each state proportional to its loss and sample states according 
to these probabilities. In more detail, the loss of a state is given by its 
TD-Error $$\delta_i = r_t +\gamma 
Q_{target}(s_{t+1},\text{argmax}_{a\in\mathcal{A}}Q(s_{t+1},a))-Q(s_t,a)\ .$$ 
Then the so called priorities are $p_i = |\delta_i| + \epsilon$ with some 
small constant to ensure that the priority is never zero (which would 
correspond to not being sampled at all).  To get to a probability, one has to 
norm these priorities to something between zero and one
\begin{equation} 
	P(i) = \frac{p_i^\alpha}{\sum_k p_k^\alpha}\ ,
	\label{eq:alpha}
\end{equation}
where $k$ runs over all states. The exponents $\alpha$ determine the level of 
prioritization, meaning that small values lead to states being sampled more 
equally, whereas high values  lead to a more loss dependent sampling. This 
method introduces an over-sampling and bias with respect to a uniform sampling. 
To correct for this bias by a small amount, importance sampling is used. For 
this purpose, one has to introduce weights 
\begin{equation} 
	w_i = \left(\frac{1}{N}\cdot\frac{1}{P(i)}\right)^\beta\ ,
	\label{eq:beta}
\end{equation}
where $N$ is the buffer size meaning the number of saved states to sample from 
and $\beta$ is a coefficient fulfilling the same role as $\alpha$ before. These 
weights are multiplied by $\delta_i$ and then used in the gradient updates of 
the network
\begin{equation}
	\Delta\theta = \alpha~ w_i\delta_i~\nabla_\theta Q^\pi_\theta(s,a)\ .
\end{equation} 
This ensures, that updates coming from high probability states don't have too 
much of an impact on the network parameters. The exponents are usually chosen 
to be $\alpha=0.1$ and $\beta=0.6$.
\subsection{Asynchronous advantage actor-critic (A3C)}
In actor-critic methods the policy (actor) as well as the Q-function (critic) 
will be approximated by a neural network 
\begin{eqnarray}
	\pi(s,a) &\simeq& \pi_\theta(s,a)\ ,\nonumber\\
	Q(s,a)&\simeq& Q_w(s,a)\ ,\nonumber
\end{eqnarray}
with the weights $\theta$ and $w$. Often, in policy approximation 
methods, a baseline is introduced to reduce the variance of these algorithms 
and 
hence to improve the learning behaviour. This can be done in actor-critic 
methods 
as well. The baseline will be subtracted to result in the following loss 
function $$\nabla_\theta\text{L}(\theta) = 
\mathbb{E}_{\pi_\theta}[\nabla_\theta\text{log}\pi_\theta(s,a)(Q^{\pi_\theta}(s,
a)-b(s))]\
 .$$ One can think of this modification as a measure of how much better our 
approximation is than is expected by some baseline $b(s)$. Since $b(s)$ does 
not depend on any action, this does not change the expectation value in the 
equation and therefore is mathematically allowed. Choosing $b(s)=V(s)$ the 
value function as a baseline is particular useful and our loss is then given by 
\begin{equation}
	\nabla_\theta\text{L}(\theta) = 	
\mathbb{E}_{\pi_\theta}[\nabla_\theta\text{log}\pi_\theta(s,a)A^{\pi_\theta}(s,
a)]\
	 , \label{eq:A2Closs}
\end{equation}
with the advantage function $A^{\pi_\theta} = 
Q^{\pi_\theta}(s,a)-V^{\pi_\theta}(s)$. In a concrete implementation of this so 
called advantage actor-critic algorithm the value function will be approximated 
by a neural network since, using the TD-algorithm, the advantage function at 
timestep $t$ is given by $A_w(s_t,a_t) = r_{t+1} +\gamma 
V_w(s_{t+1})-V_w(s_t)$. Then the critic network which approximates the value 
function will just be updated using $\nabla_wA_w(s,a).$
The last piece we need is the asynchronous advantage actor-critic (A3C) 
algorithm. Here learning and exploring is parallelized to different CPU cores. 
Each core has its own copy of the environment and of a global neural network. 
The actors then act according to their copy of the neural network and collect 
gradients by walking through their environment and computing the actor-critic 
losses. After some time (if all the different actors on the different cores 
would send their gradients at the same time this would be called synchronous 
actor-critic) an actor sends these gradients to the global network and this 
network is updated according to them. The actor then gets the new weights from 
the network and explores and computes gradients again and so on. This method 
has the advantages, that it allows for shorter training time due to the 
parallelization and for more exploration since every actor most likely explores 
different regions of the environment. To exploit this exploration benefit even 
more, the authors of \cite{mnih2016asynchronous} added an entropy 
regularization term to the policy loss \ref{eq:A2Closs}
\begin{equation}
	\nabla_\theta\text{L}(\theta) = 	
\mathbb{E}_{\pi_\theta}[\nabla_\theta\text{log}\pi_\theta(s,a)A^{\pi_\theta}(s,
a)+\beta\nabla_\theta
	 H_\theta(\pi_\theta(s))]\ ,
\end{equation}
with the entropy $ 
H_\theta(\pi_\theta(s))=-\sum_a\pi_\theta(s,a)\text{log}\pi_\theta(s,a)$.
Because this loss is to be maximized, the entropy will maximize as well and 
since the entropy measures the ''randomness'' of the policy, this will lead to 
even more exploration.
\section{RL-hyperparameters}
\label{app:hyperparameters}
\begin{figure}
	\includegraphics[width=0.49\textwidth]{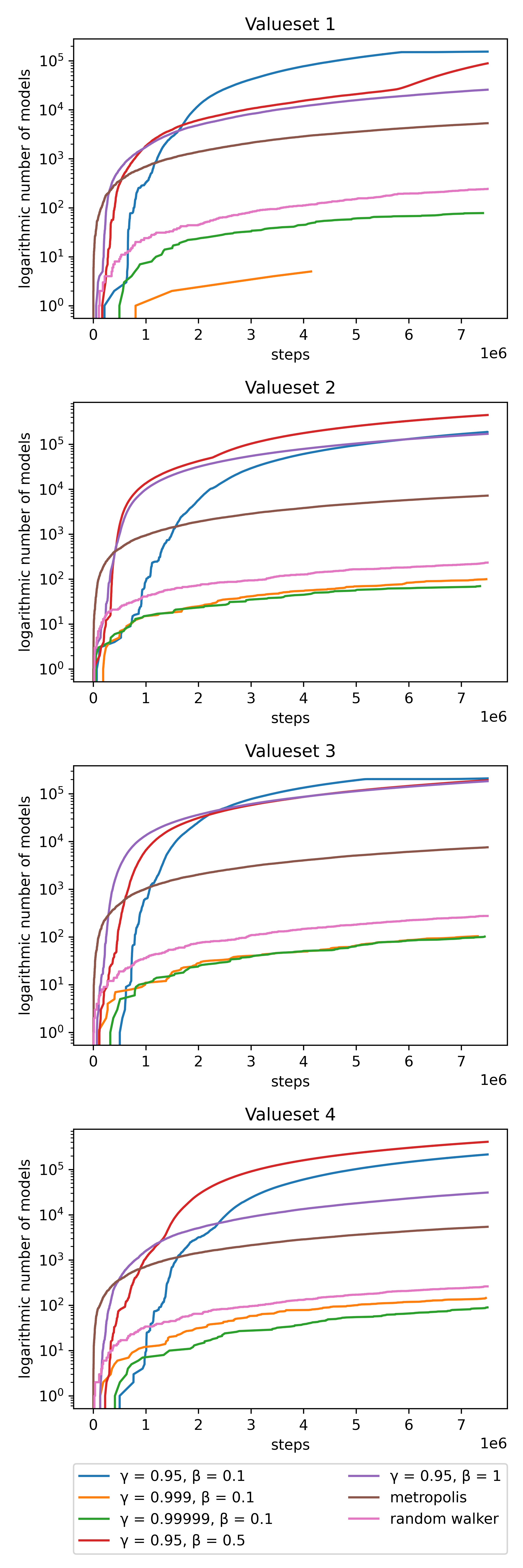}
	\includegraphics[width=0.49\textwidth]{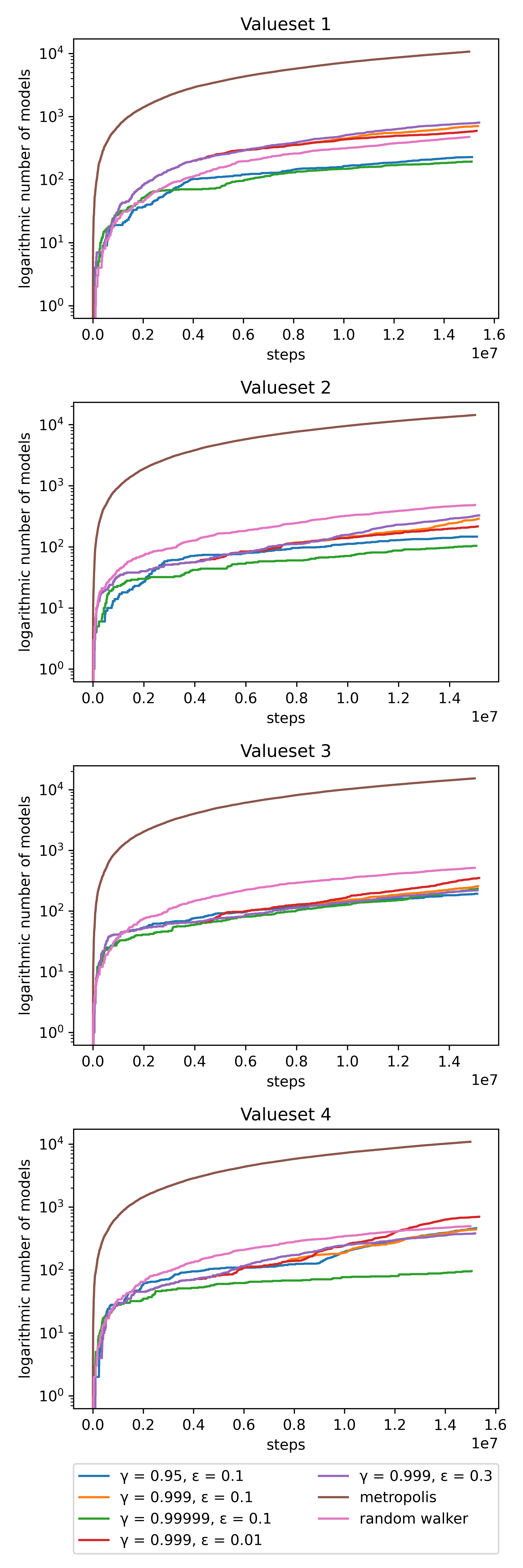}
	\caption{Results of hyperparameter scan for different reward functions 
for A3C (left) and DQN (right) experiments.}
	\label{fig:hyper_a3c}
\end{figure}

\begin{figure}
\minipage{0.49\textwidth}
	\includegraphics[width=\linewidth]{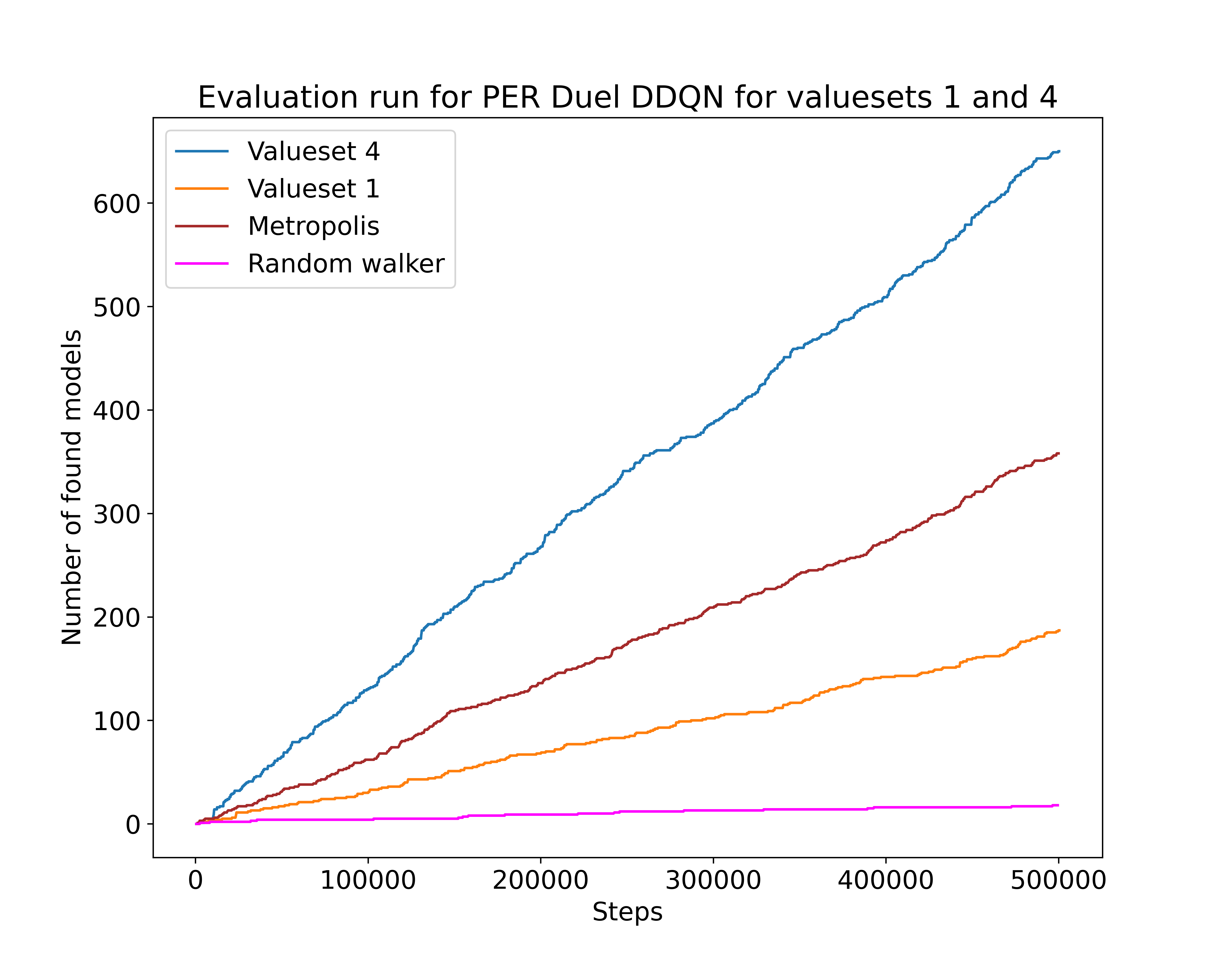}
	\caption{Results of 500000 evaluation steps for the DQN agents valuesets 
		1 and 4 and comparison to a random walker and metropolis.}
	\label{fig:dqn_evaluation_comparison}
\endminipage\hfill
\minipage{0.49\textwidth}
	\includegraphics[width=\linewidth]{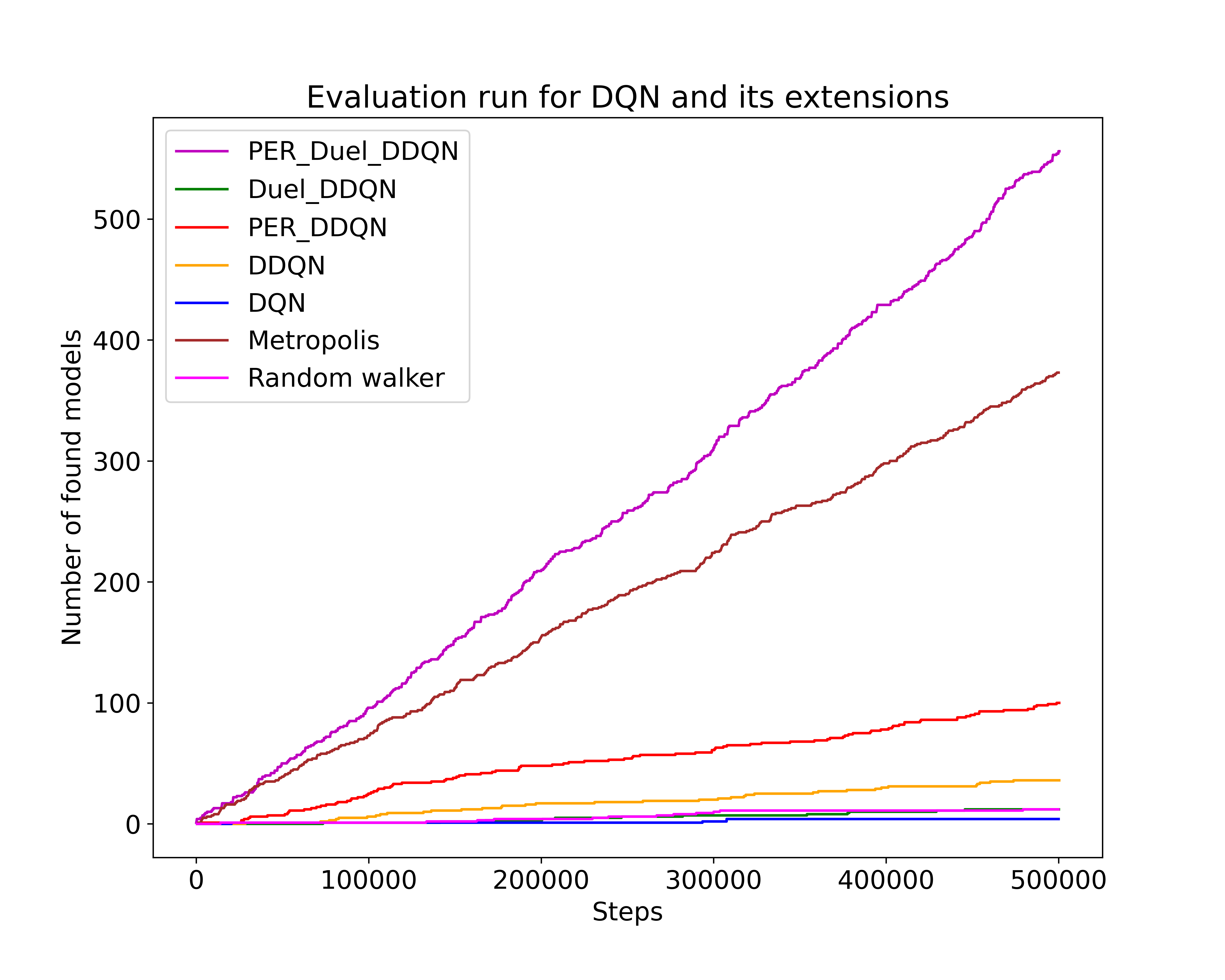}
	\caption{Comparison of DQN to its extensions: PER Duel DDQN outperforms the single extensions strongly. }
\label{fig:DQNvariants_eval}
\endminipage\hfill
\end{figure}

This appendix provides an overview of our hyperparameter searches we have 
performed.
This is necessary, because the choice of hyperparameters can make the 
difference 
between a well learning and a non working agent as we will see in the case of 
DQN. All of the experiments shown here are done on the conifold environment 
with 
flux initialization interval $[-3,3]$. For this we will consider four value 
sets with different reward functions as shown in Table \ref{table}:
\begin{table}[h]
	\centering
	\begin{tabular}{|l|l|l|l|l|l|}
		\hline
		& $N_s $ & $\sigma_s$ & $r_s$ & $\sigma_t$ & $r_t$\\
		\hline
		valueset 1  & $3$ & $10000$ & $2000$ & $0.2$ & $-5$\\
		valueset 2  & $ 5$ & $50000$ & $7000$ & $0.2$ & $-3$\\
		valueset 3  & $20$ & $100000$ & $10^6$ & $0.0005$ & $-3$\\
		valueset 4  & $3$  & $10000$ & $7000$ & $0.2$ & $-3$\\ 	
		\hline
	\end{tabular}
	\caption{Different sets of reward function hyperparameters}
	\label{table}
\end{table}
\\
For each valueset the other reward function hyperparameters are set as:
\begin{equation}
	r_g = 10,\  \sigma_g=2,\ N_g=0.5,\ N_t = 1\ .
\end{equation}
In the experiments shown in this section we will have a look at the influence of 
$\gamma$, which defines how much future rewards are valued compared to more 
short term rewards. Also, we wanted to try out different exploration rates which 
corresponds to $\beta$ in the case of A3C and to $\epsilon$ in the case of DQN. 
The value of $\epsilon$ specifies the probability, with which the agent chooses 
a random action. So a high $\epsilon$ means, that the agent often chooses a 
random action and hence explores more of the environment than he would by only 
following the action suggested by the neural network. On the other hand, the 
value of $\beta$ influences the entropy term in Equation~\eqref{eq:A2Closs} so a high value 
here leads to the agent paying more attention to maximising the entropy and 
hence leads to more exploration as well. The different values tried here are:
\begin{itemize}
	\item \textbf{A3C:} $\gamma\in\{0.95,0.999,0.99999\}, \ 
\beta\in\{0.1,0.5,1\}$
	\item \textbf{DQN:} $\gamma\in\{0.95,0.999,0.99999\}, \ 
\epsilon\in\{0.01,0.1,0.3\}$
\end{itemize}
Each experiment trains the  A3C agent for $7.5\times 10^6$ steps and the DQN 
agent for $15\times 10^6$ steps. The agent is reset every 500 steps or if he 
found a good vacuum. The results are always given in terms of the number of good 
vacua the agent found during training. Results for the A3C experiments are given on the left side 
in Figure \ref{fig:hyper_a3c}.
In all experiments, a small $\gamma$ value performs best. That means it is more 
beneficial for the agent to prefer short term rewards, so just increasing the 
distance reward step by step instead of focusing on hitting a good vacuum 
somewhere in the future. On the other hand the $\beta$ value does not have as 
much of an impact. A higher value there leads to a faster learning. However, the 
small $\beta$ always catches up and for valueset 1 and 3 even beats the others. 
So we conclude, that a higher $\beta$ which corresponds to more exploration, 
leads to a faster but less successful learning. Concerning the valueset, in all 
three cases the agent finds around 150000 not necessarily distinct models. 
However, valueset 3 gives the highest amount and hence we choose to work with 
that one.  So our best working agent uses valueset 3 with hyperparameters 
$\gamma=0.95$ and $\beta=0.1$. In all cases the metropolis algorithm finds order 
$\mathcal{O}(20)$ less models then our agent, while a random walker finds order 
$\mathcal{O}(600)$ less.\\
Next, the results for the DQN experiments are given on the right side in Figure 
\ref{fig:hyper_a3c}. The first thing to note is, that a high $\gamma$ performs 
worst again. Different to A3C though it is the middle value for $\gamma = 0.999$ 
that works best. Concerning the exploration parameter $\epsilon$, the 
performance seems to depend on the valueset. For valuesets 1 and 2 a higher 
$\epsilon$ and hence taking a random action more often works better. For 
valuesets 3 and 4 however exploiting works best. If we now focus on the total 
number of found models, we see that not a single agent beats metropolis and only 
for valueset 1 and 4 some manage to beat a random walker. Also, the slope of the 
model curve does only change for valueset 4 and hence only there some kind of 
learning behaviour is present. The best agent for valueset 1 finds roughly 800 
models while for valueset 4 he finds 700. In both cases that is a factor of 1.5 
more then for a random walker. However, metropolis finds $\mathcal{O}(20)$ more. 
All of these insights are for the models found during training. If we look at 
some evaluation runs of trained agents and compare those to metropolis and a 
random walker, the results are given in Figure 
\ref{fig:dqn_evaluation_comparison}. There we see, that only for valueset 4, the 
agent manages to beat metropolis by a factor of 2. Hence we choose valueset 4 as 
our best working reward function for DQN.
\begin{figure}
\minipage{0.49\textwidth}
	\includegraphics[width=\linewidth]{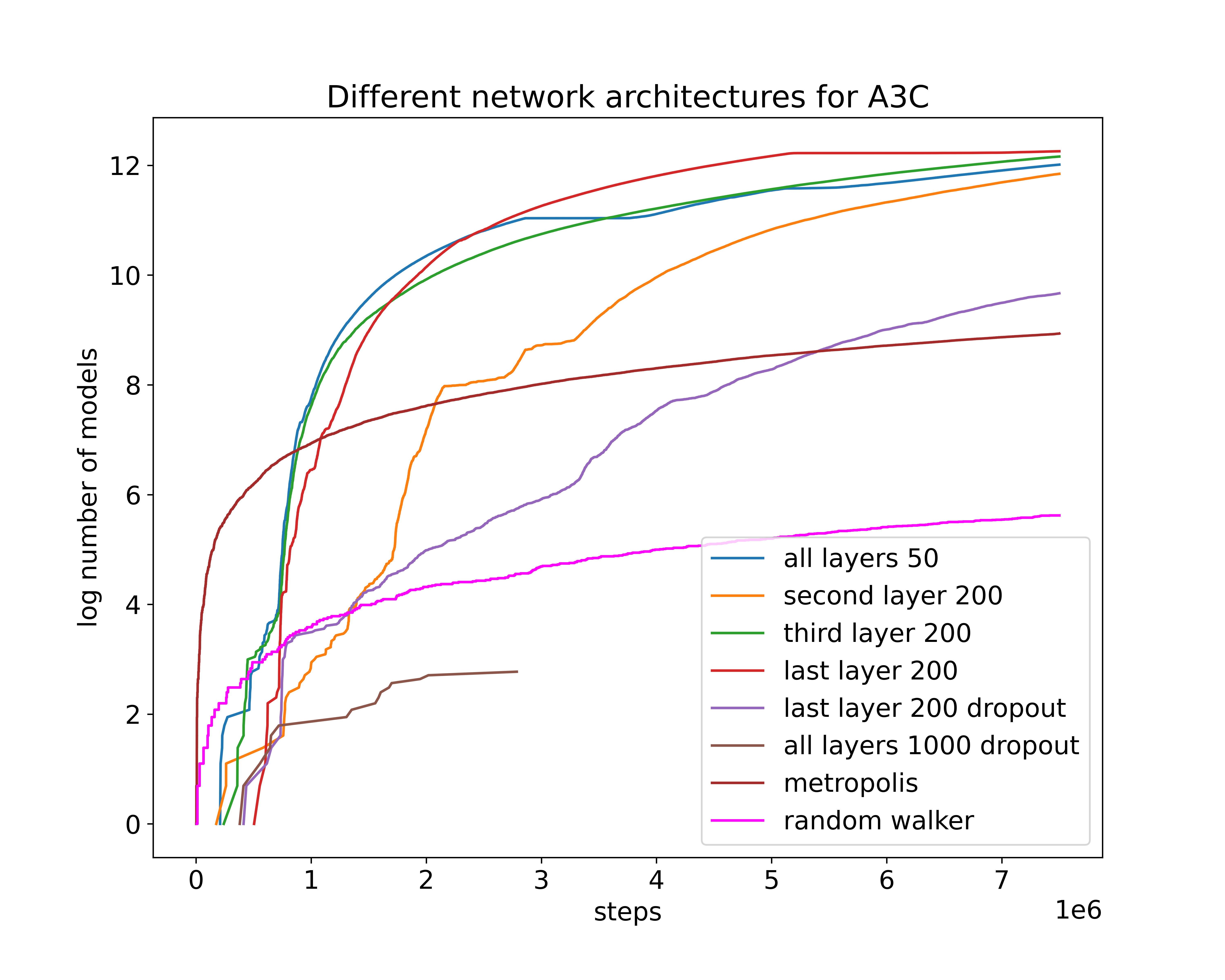}
	\caption{Logarithmic number of found models for A3C experiments with different model complexities.}
	\label{fig:model_complexity}
\endminipage\hfill
\minipage{0.49\textwidth}
	\includegraphics[width=\linewidth]{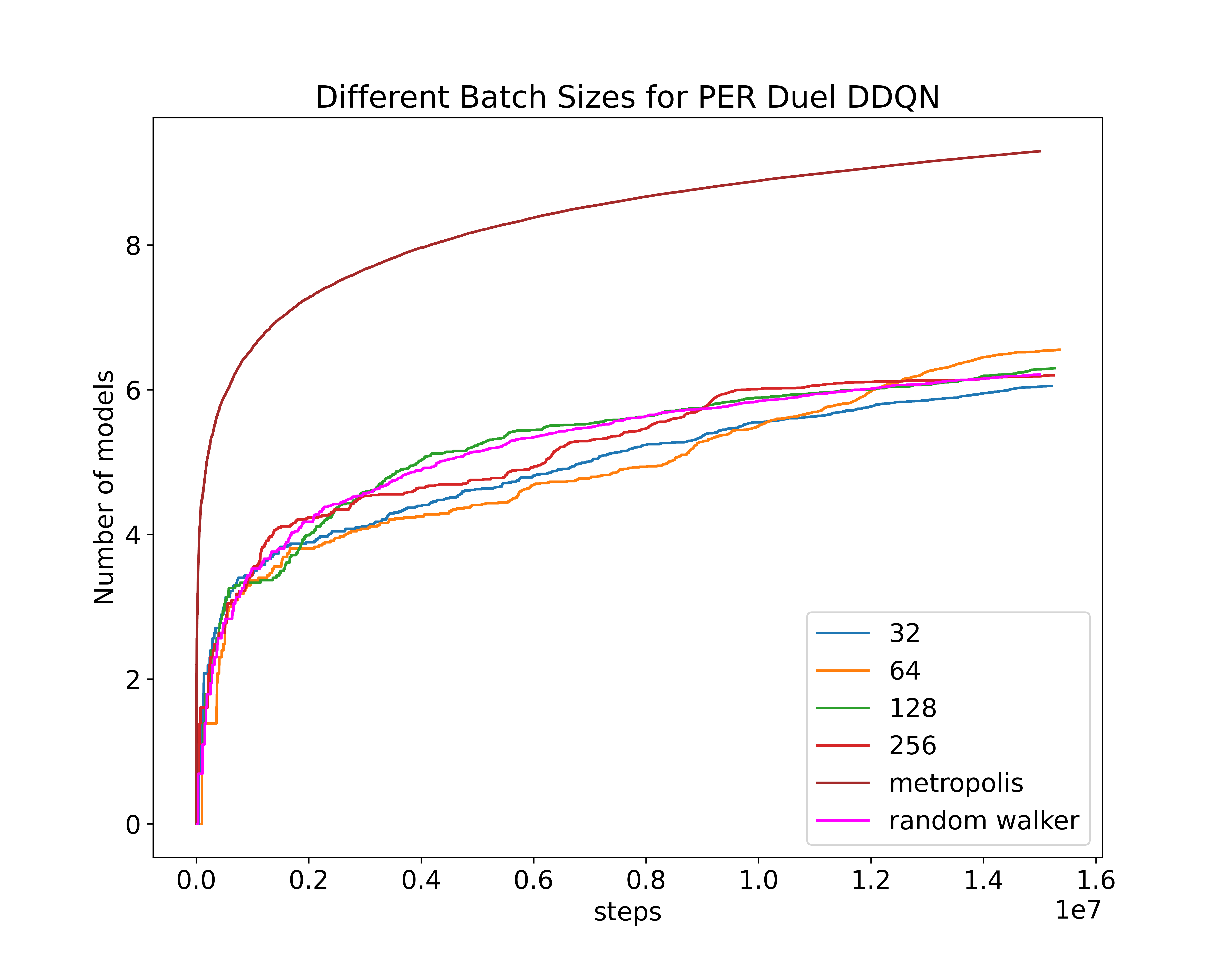}
	\caption{Logarithmic number of found models for DQN experiments with different batch sizes.}
	\label{fig:batch_size}
\endminipage\hfill
\end{figure}
\newpage
\subsection*{Model complexity}
Here we want to study the influence of different neural network architectures. 
For that purpose, we will look at an A3C agent with optimal hyperparameters and 
valueset, i.e. $\gamma=0.95$, $\beta=0.1$ and valueset 3 respectively. The 
network will always be dense and consist of 4 hidden layers. However, the size 
of these layers will be varied in the following way:
\begin{table}[h]
	\centering
	\begin{tabular}{|l|l|l|l|l|}
		\hline
		Layer number & 1 & 2 & 3 & 4 \\
		\hline
		Layer size & 50 & 50 & 50 & 50 \\
		& 50 & 200 & 50 & 50 \\
		& 50 & 50 & 200 & 50 \\
		& 50 & 50 & 50 & 200 \\
		& 1000 & 1000 & 1000 & 1000 \\				
		\hline
	\end{tabular}
	\caption{Different sets of reward function hyperparameters.}
\end{table}
Also, we tried to include dropouts with probability $p=0.5$. In all these 
experiments the agent was run for $7.5\cdot10^6$ steps or 120 hours whatever 
occurred first. The results in terms of logarithmic number of found models is 
shown in Figure \ref{fig:model_complexity}. First, we notice that the run with 
layer size 1000 was stopped early since it reached the 120 hours time mark. So 
even if it might perform better at late times, it is not suitable for our 
purpose. Dropout does not perform well either. All other architectures do not 
differ much to each other, but the red curve reaches the most models and an 
equally fast learning behaviour, so we choose to work with that one. We did not 
try convolutional architectures since our problem is not translation invariant. 
 
\subsubsection*{Batch size}
Next we study the influence of the batch size on the DQN agent Figure 
\ref{fig:batch_size}. 
As we can see, only for batch size 64 the curve gets steeper at around 
$9\cdot10^6$ steps and hence only there the agent shows some learning behaviour. 
Therefore we choose to work with that batch size.   
\subsection*{DQN variants}
\begin{figure}\begin{center}
	\includegraphics[width=0.5\textwidth]{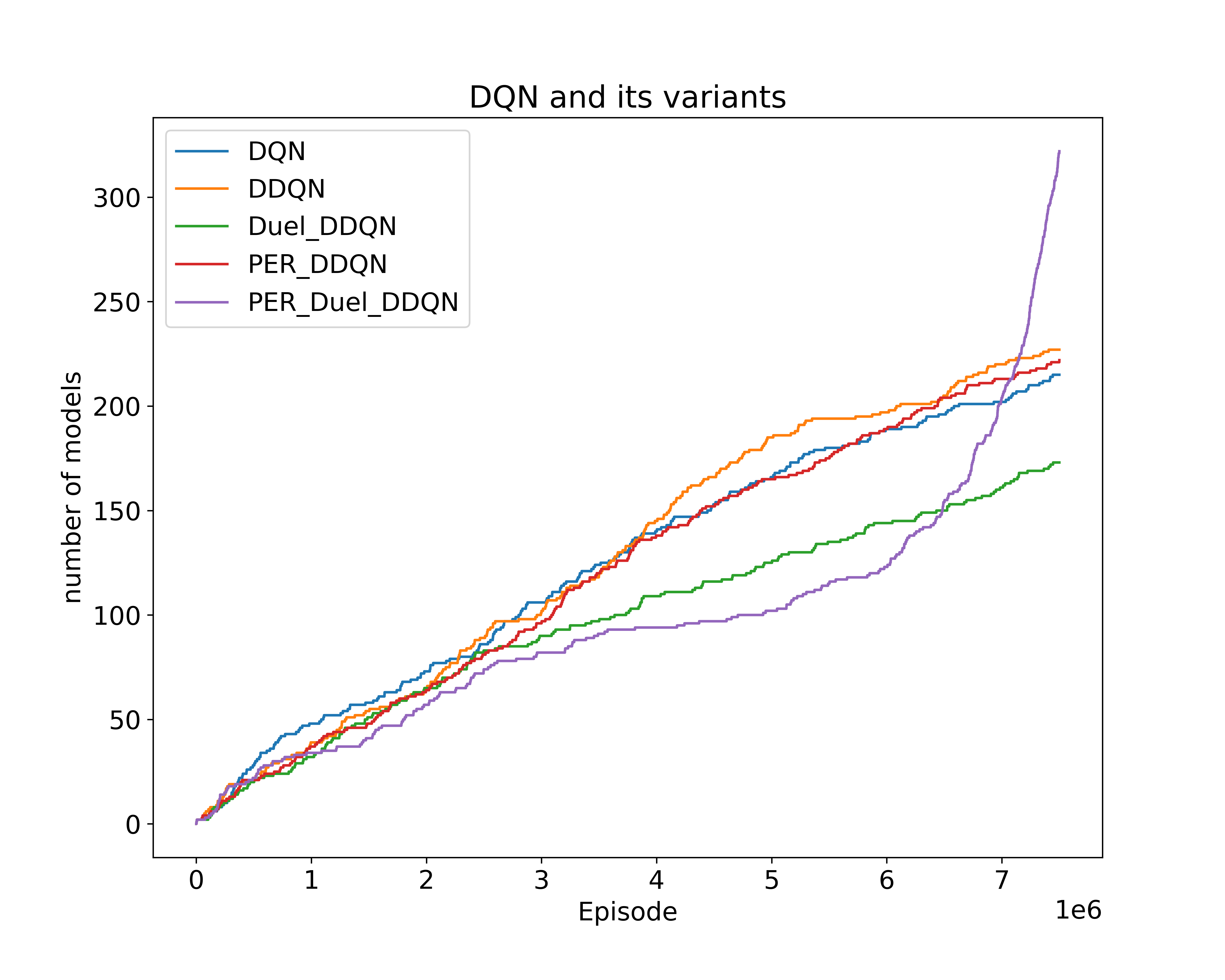}\end{center}
	\caption{Comparison of DQN to its extensions: PER Duel DDQN contains all 
other shown extensions to DQN and adds prioritized experience replay. Hence, it 
outperforms the single extensions strongly.}
	\label{fig:DQNvariants}
\end{figure} 
The next plot Figure \ref{fig:DQNvariants} shows the comparison of DQN to its 
different extensions. All hyperparameters are put to default values and valueset 
4 was used. The agents were trained for $7.5\cdot10^6$ steps and reset after 
they found a model or 500 steps. One can see, that PER Dueling DDQN performs 
best. The same holds for the comparison on the level of evaluation Figure 
\ref{fig:DQNvariants_eval}.  The reason for this might be the nature of 
prioritized experience replay. Since in our environment, states fulfilling all 
of our conditions are rare, the agent won't see them very often during training 
and hence can not learn much from them. Due to PER though, they can be sampled 
more often leading to a better performance.

%\newpage
\section{Torus environment}
\label{app:torusenvironments}
In this appendix, we will present more details about our experiments on the 
torus. We looked for specific values of the superpotential $|W_0|$ and the 
string coupling $g_s$. 
\subsection*{Superpotential search}
Here we will look for solutions with $0<|W_0|<10$. Of course the
dilaton~\eqref{eq:fdomain} and tadpole condition~\eqref{eq:tadpole condition} also have to 
be satisfied. Our algorithm will calculate the dilaton for all real solutions 
of Equation~\eqref{eq:cubic_torus}. The superpotantial then will only be calculated for 
the values of $\tau$ for which the dilaton lies in the fundamental domain. The 
structure of the environment are the same as in chapter 3. The neural network 
contains again 4 dense layers with 50 neurons for the first three and 200 for 
the last layer. The distance rewards (punishments) are given by
\begin{eqnarray}
	r_g&=& 
	\begin{cases}
		N_g\cdot 
		
\text{exp}\left(-\left(\frac{|\text{Re}(\phi)|-0.5}{\sigma_g}
\right)^2\right)\cdot
		 r_{\rm gauge} \ &\text{if Im}(\phi)>1\ ,\\
		N_g\cdot 
		
\text{exp}\left(-\left(\frac{|\text{Re}(\phi)|-0.5}{\sigma_g}
\right)^2\right)\cdot

\text{exp}\left(-\left(\frac{\text{Im}(\phi)-1}{\sigma_g}\right)^2\right)\cdot
		 r_{\rm gauge}\ &\text{else}\ ,
	\end{cases}\\
	r_{t} &=&
	\begin{cases}
		N_t\cdot\text{tanh}\left(\sigma_{t_1}\left( 
		N_\text{flux}-0\right)\right)\cdot r_\text{tad} \ \text{if}\  
		N_\text{flux}<0 \ ,\\
		N_t\cdot\text{tanh}\left(\sigma_{t_2}\left( 
		N_\text{flux}-0\right)\right)\cdot r_\text{tad} \ \text{if}\ 
		N_\text{flux}>16 \ ,  
	\end{cases}\\
	r_s &=& 
	N_s\cdot\text{exp}\left(-\left(\frac{|W_0|-0}{\sigma_s}\right)^2\right)\ 
,
\end{eqnarray}   
with the constants
\begin{eqnarray}
	N_g = 0.5\ , &\quad \sigma_g = 2\ , &\quad r_\text{gauge}= 10 \ , \\
	N_t = 1\ , &\quad \sigma_{t_1} = 0.003\ , &\quad \sigma_{t_2} = 0.0005\ 
, 
	\quad r_\text{tad} = -5\\
	N_s = 20\ ,&\quad \sigma_s = 3000\ , &\quad r_\text{sup} = 100000\ .
\end{eqnarray}
As RL agent, we used the best two from the conifold experiments, namely an A3C 
and a prioritized experience replay duel DDQN. They are again compared to a 
random walker and a metropolis algorithm. The hyperparameters are chosen the 
same way as for the conifold. The agent was trained for $7.5\cdot 10^6$ steps 
and reset after he found a model or after 500 steps.

\bibliography{references}{}
	\bibliographystyle{JHEP}

\end{document}